
\font\gross=cmbx10  scaled\magstep2
\font\mittel=cmbx10 scaled\magstep1
\def\gsim{\mathrel{\raise.3ex\hbox{$>$\kern- .75em
                      \lower1ex\hbox{$\sim$}}}}
\def\square{\kern 1pt
\vbox{\hrule height 0.6pt\hbox{\vrule width 0.6pt \hskip 3pt
\vbox{\vskip 6pt}\hskip 3pt\vrule width 0.6pt} \hrule height 0.6pt}\kern 1pt }

\def\sla{\raise.15ex\hbox{$/$}\kern-.72em}


\parskip=\medskipamount
\overfullrule=0pt
\raggedbottom
\def\normalparindent{24pt}
\newif\ifdraft \draftfalse

\nopagenumbers
\footline={\ifnum\pageno=1 {\ifdraft
{\hfil\rm Draft \number\day -\number\month -\number\year}
\else{\hfil}\fi}
\else{\hfil\rm\folio\hfil}\fi}
\def\endpage{\vfill\eject}
\def\beginlinemode{\endmode\begingroup\parskip=0pt
\obeylines\def\\{\par}\def\endmode{\par\endgroup}}
\def\beginparmode{\endmode\begingroup \def\endmode{\par\endgroup}}
\let\endmode=\par
\def\raggedcenter{
                  \leftskip=2em plus 6em \rightskip=\leftskip
                  \parindent=0pt \parfillskip=0pt \spaceskip=.3333em
                  \xspaceskip=.5em\pretolerance=9999 \tolerance=9999
                  \hyphenpenalty=9999 \exhyphenpenalty=9999 }
\def\\{\cr}
\let\rawfootnote=\footnote\def\footnote#1#2{{\parindent=0pt\parskip=0pt
        \rawfootnote{#1}{#2\hfill\vrule height 0pt depth 6pt width 0pt}}}
\def\title{\null\vskip 3pt plus 0.2fill\beginlinemode\raggedcenter\gross}
\def\author{\vskip 3pt plus 0.2fill \beginlinemode\raggedcenter}

\def\abstract{\vskip 3pt plus 0.3fill \beginparmode{\noindent
{\mittel Abstract}:~}  }
\def\endtitlepage{\endpage\body}
\def\body{\beginparmode\parindent=\normalparindent}
\def\head#1{\par\goodbreak{\immediate\write16{#1}
      \vskip 0.4cm{\noindent\gross #1}\par}\nobreak\nobreak\nobreak\nobreak}
\def\subhead#1{\par\goodbreak{\immediate\write16{#1}
           {\noindent\mittel #1}\par}\nobreak\nobreak}

\gdef\refto#1{\ifprdstyle  $^{\[#1] }$ \else
              \ifwsstyle$^{\[#1]}$  \else
              \ifannpstyle $~[\[#1] ]$ \else
              \ifplbstyle  $~[\[#1] ]$ \else
                                         $^{[\[#1] ]}$\fi\fi\fi\fi}
\gdef\refis#1{\ifprdstyle \item{~$^{#1}$}\else
              \ifwsstyle \item{#1.} \else
              \ifplbstyle\item{~[#1]} \else
              \ifannpstyle \item{#1.} \else
                              \item{#1.\ }\fi\fi\fi\fi }
\def\finalcite{\citeall\ref\citeall\Ref}
\newif\ifannpstyle
\newif\ifprdstyle
\newif\ifplbstyle
\newif\ifwsstyle

\gdef\refto#1{\ifprdstyle  $^{\[#1] }$ \else
              \ifwsstyle$^{\[#1]}$  \else
              \ifannpstyle $~[\[#1] ]$ \else
              \ifplbstyle  $~[\[#1] ]$ \else
                                         $^{[\[#1] ]}$\fi\fi\fi\fi}
\gdef\refis#1{\ifprdstyle \item{~$^{#1}$}\else
              \ifwsstyle \item{#1.} \else
              \ifplbstyle\item{~[#1]} \else
              \ifannpstyle \item{#1.} \else
                              \item{#1.\ }\fi\fi\fi\fi }
\gdef\journal#1,#2,#3,#4.{
           \ifprdstyle {#1~}{\bf #2}, #3 (#4).\else
           \ifwsstyle {\it #1~}{\bf #2~} (#4) #3.\else
           \ifplbstyle {#1~}{#2~} (#4) #3.\else
           \ifannpstyle {\sl #1~}{\bf #2~} (#4), #3.\else
                       {\sl #1~}{\bf #2}, #3 (#4)\fi\fi\fi\fi}

\def\ref#1{Ref.~#1}
\def\Ref#1{Ref.~#1}
\def\cite#1{{#1}}\def\[#1]{\cite{#1}}

\def\(#1){(\call{#1})}
\def\call#1{{#1}}\def\taghead#1{{#1}}

\def\references{\head{References}\beginparmode\frenchspacing\parskip=0pt}
\def\endreferences{\body}
\def\endit{\endmode\vfill\supereject}\let\endpaper=\endit


\def\prd{\journal Phys. Rev. D,}
\def\prl{\journal Phys. Rev. Lett.,}
\def\ijmpa{\journal Int. J. Mod. Phys. A,}

\def\npb{\journal Nucl. Phys. B,}

\def\plb{\journal Phys. Lett. B,}
\def\grg{\journal Gen. Relativ. Grav.,}
\def\cqg{\journal Class. Quantum Grav.,}

\def\annp{\journal Ann. Phys. (N.Y.),}


\def\barce{Universidad Aut\'onoma de Barcelona, IFAE, Grupo de F\'\i sica
Te\'orica, E-08193 Bellaterra (Barcelona), Spain. E-mail: lousto@ifae.es}

\def\demdir{Observatoire de Paris DEMIRM, UA 336 Laboratoire
associ\' e au CNRS,\\ 61, Av. de l'Observatoire, 75014 Paris, France.}

\def\iafedir{Permanent Address: Instituto de Astronom\'\i a y F\'\i sica del
Espacio, Casilla de Correo 67 -\\ Sucursal 28, 1428 Buenos Aires, Argentina.
E-mail: lousto@iafe.edu.ar}

\def\cee{This work was partially supported by the Directorate General for
Science, Research and Development of the Commission of the European
Communities.}

\def\guita{C.O.L. was financially supported by the Direcci\'on General de
Investigaci\'on Cient\'\i fica y T\'ecnica of the Ministerio de Educaci\'on
y Ciencia de Espa\~na and CICYT AEN93-0474.}

\catcode`@=11
\newcount\r@fcount \r@fcount=0\newcount\r@fcurr
\immediate\newwrite\reffile\newif\ifr@ffile\r@ffilefalse
\def\w@rnwrite#1{\ifr@ffile\immediate\write\reffile{#1}\fi\message{#1}}
\def\writer@f#1>>{}
\def\referencefile{\r@ffiletrue\immediate\openout\reffile=\jobname.ref%
  \def\writer@f##1>>{\ifr@ffile\immediate\write\reffile%
    {\noexpand\refis{##1} = \csname r@fnum##1\endcsname = %
     \expandafter\expandafter\expandafter\strip@t\expandafter%
     \meaning\csname r@ftext\csname r@fnum##1\endcsname\endcsname}\fi}%
  \def\strip@t##1>>{}}

\def\citeall#1{\xdef#1##1{#1{\noexpand\cite{##1}}}}
\def\cite#1{\each@rg\citer@nge{#1}}
\def\each@rg#1#2{{\let\thecsname=#1\expandafter\first@rg#2,\end,}}
\def\first@rg#1,{\thecsname{#1}\apply@rg}
\def\apply@rg#1,{\ifx\end#1\let\next=\relax%
\else,\thecsname{#1}\let\next=\apply@rg\fi\next}%
\def\citer@nge#1{\citedor@nge#1-\end-}
\def\citer@ngeat#1\end-{#1}
\def\citedor@nge#1-#2-{\ifx\end#2\r@featspace#1
  \else\citel@@p{#1}{#2}\citer@ngeat\fi}
\def\citel@@p#1#2{\ifnum#1>#2{\errmessage{Reference range #1-#2\space is bad.}
    \errhelp{If you cite a series of references by the notation M-N, then M and
    N must be integers, and N must be greater than or equal to M.}}\else%
{\count0=#1\count1=#2\advance\count1 by1\relax\expandafter\r@fcite\the\count0,%
  \loop\advance\count0 by1\relax
    \ifnum\count0<\count1,\expandafter\r@fcite\the\count0,%
  \repeat}\fi}
\def\r@featspace#1#2 {\r@fcite#1#2,}    \def\r@fcite#1,{\ifuncit@d{#1}
    \expandafter\gdef\csname r@ftext\number\r@fcount\endcsname%
    {\message{Reference #1 to be supplied.}\writer@f#1>>#1 to be supplied.\par
     }\fi\csname r@fnum#1\endcsname}
\def\ifuncit@d#1{\expandafter\ifx\csname r@fnum#1\endcsname\relax%
\global\advance\r@fcount by1%
\expandafter\xdef\csname r@fnum#1\endcsname{\number\r@fcount}}
\let\r@fis=\refis   \def\refis#1#2#3\par{\ifuncit@d{#1}%
    \w@rnwrite{Reference #1=\number\r@fcount\space is not cited up to now.}\fi%
  \expandafter\gdef\csname r@ftext\csname r@fnum#1\endcsname\endcsname%
  {\writer@f#1>>#2#3\par}}
\def\r@ferr{\endreferences\errmessage{I was expecting to see
\noexpand\endreferences before now;  I have inserted it here.}}
\let\r@ferences=\references
\def\references{\r@ferences\def\endmode{\r@ferr\par\endgroup}}
\let\endr@ferences=\endreferences
\def\endreferences{\r@fcurr=0{\loop\ifnum\r@fcurr<\r@fcount
    \advance\r@fcurr by 1\relax\expandafter\r@fis\expandafter{\number\r@fcurr}%
    \csname r@ftext\number\r@fcurr\endcsname%
  \repeat}\gdef\r@ferr{}\endr@ferences}
\let\r@fend=\endpaper\gdef\endpaper{\ifr@ffile
\immediate\write16{Cross References written on []\jobname.REF.}\fi\r@fend}
\catcode`@=12
\finalcite
\catcode`@=11
\newcount\tagnumber\tagnumber=0
\immediate\newwrite\eqnfile\newif\if@qnfile\@qnfilefalse
\def\write@qn#1{}\def\writenew@qn#1{}
\def\w@rnwrite#1{\write@qn{#1}\message{#1}}
\def\@rrwrite#1{\write@qn{#1}\errmessage{#1}}
\def\taghead#1{\gdef\t@ghead{#1}\global\tagnumber=0}
\def\t@ghead{}\expandafter\def\csname @qnnum-3\endcsname
  {{\t@ghead\advance\tagnumber by -3\relax\number\tagnumber}}
\expandafter\def\csname @qnnum-2\endcsname
  {{\t@ghead\advance\tagnumber by -2\relax\number\tagnumber}}
\expandafter\def\csname @qnnum-1\endcsname
  {{\t@ghead\advance\tagnumber by -1\relax\number\tagnumber}}
\expandafter\def\csname @qnnum0\endcsname
  {\t@ghead\number\tagnumber}
\expandafter\def\csname @qnnum+1\endcsname
  {{\t@ghead\advance\tagnumber by 1\relax\number\tagnumber}}
\expandafter\def\csname @qnnum+2\endcsname
  {{\t@ghead\advance\tagnumber by 2\relax\number\tagnumber}}
\expandafter\def\csname @qnnum+3\endcsname
  {{\t@ghead\advance\tagnumber by 3\relax\number\tagnumber}}
\def\equationfile{\@qnfiletrue\immediate\openout\eqnfile=\jobname.eqn%
  \def\write@qn##1{\if@qnfile\immediate\write\eqnfile{##1}\fi}
  \def\writenew@qn##1{\if@qnfile\immediate\write\eqnfile
    {\noexpand\tag{##1} = (\t@ghead\number\tagnumber)}\fi}}
\def\callall#1{\xdef#1##1{#1{\noexpand\call{##1}}}}
\def\call#1{\each@rg\callr@nge{#1}}
\def\each@rg#1#2{{\let\thecsname=#1\expandafter\first@rg#2,\end,}}
\def\first@rg#1,{\thecsname{#1}\apply@rg}
\def\apply@rg#1,{\ifx\end#1\let\next=\relax%
\else,\thecsname{#1}\let\next=\apply@rg\fi\next}
\def\callr@nge#1{\calldor@nge#1-\end-}\def\callr@ngeat#1\end-{#1}
\def\calldor@nge#1-#2-{\ifx\end#2\@qneatspace#1 %
  \else\calll@@p{#1}{#2}\callr@ngeat\fi}
\def\calll@@p#1#2{\ifnum#1>#2{\@rrwrite{Equation range #1-#2\space is bad.}
\errhelp{If you call a series of equations by the notation M-N, then M and
N must be integers, and N must be greater than or equal to M.}}\else%
{\count0=#1\count1=#2\advance\count1 by1\relax\expandafter\@qncall\the\count0,%
  \loop\advance\count0 by1\relax%
    \ifnum\count0<\count1,\expandafter\@qncall\the\count0,  \repeat}\fi}
\def\@qneatspace#1#2 {\@qncall#1#2,}
\def\@qncall#1,{\ifunc@lled{#1}{\def\next{#1}\ifx\next\empty\else
  \w@rnwrite{Equation number \noexpand\(>>#1<<) has not been defined yet.}
  >>#1<<\fi}\else\csname @qnnum#1\endcsname\fi}
\let\eqnono=\eqno\def\eqno(#1){\tag#1}\def\tag#1$${\eqnono(\displayt@g#1 )$$}
\def\aligntag#1\endaligntag  $${\gdef\tag##1\\{&(##1 )\cr}\eqalignno{#1\\}$$
  \gdef\tag##1$${\eqnono(\displayt@g##1 )$$}}
\def\eqalignno#1{\displ@y \tabskip\centering
  \halign to\displaywidth{\hfil$\displaystyle{##}$\tabskip\z@skip
    &$\displaystyle{{}##}$\hfil\tabskip\centering
    &\llap{$\displayt@gpar##$}\tabskip\z@skip\crcr
    #1\crcr}}
\def\displayt@gpar(#1){(\displayt@g#1 )}
\def\displayt@g#1 {\rm\ifunc@lled{#1}\global\advance\tagnumber by1
        {\def\next{#1}\ifx\next\empty\else\expandafter
        \xdef\csname @qnnum#1\endcsname{\t@ghead\number\tagnumber}\fi}%
  \writenew@qn{#1}\t@ghead\number\tagnumber\else
        {\edef\next{\t@ghead\number\tagnumber}%
        \expandafter\ifx\csname @qnnum#1\endcsname\next\else
        \w@rnwrite{Equation \noexpand\tag{#1} is a duplicate number.}\fi}%
  \csname @qnnum#1\endcsname\fi}
\def\eqnoa(#1){\global\advance\tagnumber by1\multitag{#1}{a}}
\def\eqnob(#1){\multitag{#1}{b}}
\def\eqnoc(#1){\multitag{#1}{c}}
\def\eqnod(#1){\multitag{#1}{d}}
\def\multitag#1#2$${\eqnono(\multidisplayt@g{#1}{#2} )$$}
\def\multidisplayt@g#1#2 {\rm\ifunc@lled{#1}
        {\def\next{#1}\ifx\next\empty\else\expandafter
        \xdef\csname @qnnum#1\endcsname{\t@ghead\number\tagnumber b}\fi}%
  \writenew@qn{#1}\t@ghead\number\tagnumber #2\else
        {\edef\next{\t@ghead\number\tagnumber #2}%
        \expandafter\ifx\csname @qnnum#1\endcsname\next\else
    \w@rnwrite{Equation \noexpand\multitag{#1}{#2} is a duplicate number.}\fi}%
  \csname @qnnum#1\endcsname\fi}
\def\ifunc@lled#1{\expandafter\ifx\csname @qnnum#1\endcsname\relax}
\let\@qnend=\end\gdef\end{\if@qnfile
\immediate\write16{Equation numbers written on []\jobname.EQN.}\fi\@qnend}


\catcode`\|=13

\catcode`\;=13

\def\table{
     \begingroup
       \catcode`\|=13
       \catcode`\;=13
       \def|{&\I&&} 
       \def;{&&&}   
       \def\II{\vrule width 2.0pt}
       \def\I{\vrule}
       \def~{\hfill}
       \def\\{\strut\cr}
       \def\>{\hbox{\vbox to 14pt{}\vtop to 9pt{}}\cr}
       \def\space##1{\vbox to ##1pt{}}
       \def\----{\noalign{\hrule}\cr}
       \def\===={\noalign{\hrule height 2.0pt}\cr}
       \def\thinline{&\multispan3\hrulefill}
       \def\thickline{&\multispan3{\leaders\hrule height 2.0pt\hfill}}
       \def\noline{&\multispan2&}
        \def\makeline##1{##1\cr}
       \def\[{\II&&} 
       \def\]{&\II \>} 
       \def\({&&} 
       \def\){&\\} 
     \tableaux}

\def\tableaux#1#2{\vbox{
      \tabskip=0pt
      \offinterlineskip
      \parskip=0pt
     \halign to #1{&\tabskip=0pt##& ##\tabskip=0.5em plus 4em minus0.5em
     &\hfil##\hfil\cr
      #2}}
    \endtable}

\def\endtable{\catcode`\|=12 \catcode`\;=12\endgroup}
\catcode`\|=12
\catcode`\;=12

\baselineskip=12pt
\magnification=1200
\hfill{preprint UAB-FT-353, gr-qc/9410041}
\bigskip\bigskip
\bigskip\bigskip
\parindent=0pt
\title
{SCATTERING PROCESSES

AT THE PLANCK SCALE}

\bigskip\bigskip
\author
C. O. LOUSTO\footnote{$^{a}$}{\iafedir}
\barce
\bigskip
and
\bigskip
N. S\'ANCHEZ
\demdir
\bigskip\bigskip

\abstract
The ultrarelativistic limit of the Kerr - Newman geometry is studied in detail
We find the corresponding gravitational shock wave background associated with
this limit. Interestingly, this allows us to find the source of the Kerr -
Newman geometry in the ultrarelativistic regime. We study the scattering of
scalar fields in the gravitational shock wave geometries,
 and discuss the presence of the poles
$iGs=n=0,1,2,...$, already present in the
Aichelburg - Sexl metric. We compare this with the scattering by
ultrarelativistic extended sources, for which such poles do not appear and
with the scattering of fundamental strings.

We also study planckian energy string collisions in flat spacetime as the
scattering of a string in the effective curved background produced by the
 others as the impact parameter $b$ decreases.  We find the
 effective  energy density distribution
$\sigma(\rho)\sim\exp\{-\rho^2/\Delta^2\}$, generated by these collisions.
Two different regimes can be studied: intermediate
 impact parameters  $x^d\ll b\ll\sqrt{\alpha '\ln s}\cong\Delta/2\quad,$
( $x^d$ characterizing the string fluctuations) and
 large impact parameters,
$b\gg\sqrt{\alpha '\ln s}\cong\Delta/2\gg x^d~.$
The effective metric generated by these collisions
is a gravitational shock wave  of profile $f(\rho)\sim p\rho^{4-D}$ ,
 i.e. the Aichelburg - Sexl geometry for a point-like particle of
momentum $p$ for large $b$.
 For intermediate $b$,  $f(\rho)\sim q\rho^2$ ,
corresponding to an extended source of momentum $q$.
The scattering matrix in this geometry and its
 implications for the string collision process are
analysed. We show that the poles $iGs=n~,~n=0,1,2...$, characteristic of
the scattering by the A - S geometry are absent here, due to the extended
nature of the effective source.

We finally study the emergence of string instabilities in
$D$ - dimensional black
hole spacetimes (Schwarzschild and Reissner - Nordstr\"om), and De Sitter
space (in static coordinates to allow a better comparison with the black
hole case). We solve the first order string fluctuations around the center
of mass motion at spatial infinity, near the horizon and at the spacetime
singularity. We find that the time components are always well behaved in
the three regions and in the three backgrounds. The radial components are
{\it unstable}: imaginary frequencies develop in the oscillatory modes near
the horizon, and the evolution is like $(\tau-\tau_0)^{-P}$, $(P>0)$, near
the spacetime singularity, $r\to0$, where the world - sheet time
$(\tau-\tau_0)\to0$, and the proper string length grows infinitely. In the
Schwarzschild black hole, the angular components are always well - behaved,
while in the Reissner - Nordstr\"om case they develop instabilities inside
the horizon, near $r\to0$ where the repulsive effects of the charge dominate
over those of the mass. In general, whenever large enough repulsive effects
in the gravitational background are present, string instabilities develop.
In De Sitter space, all the spatial components exhibit instability. The
infalling of the string to the black hole singularity is like the motion of a
particle in a potential $\gamma(\tau-\tau_0)^{-2}$ where $\gamma$ depends on
the $D$ spacetime dimensions and string angular momentum, with $\gamma>0$
for Schwarzschild and $\gamma<0$ for Reissner - Nordstr\"om black holes.
For $(\tau-\tau_0)\to0$ the string ends trapped by the black hole singularity.
\parindent=10pt
\endtitlepage
\head{1. Introduction}

Recently, gravitational shock waves backgrounds have raised interest in the,
context of both Field Theory and
Strings\refto{H87,dVS89b,dVS89a,ACV88,GM88,AK88,V88,23,dVS90,GV91,HS94,AB94}.
Besides their role in classical gravitation, as an exact class of solutions,
these geometries are relevant at energies of the order of the Planck scale.

At energies of the order of the Planck scale, the picture of particles
propagating in flat space - time is no longer valid, and one must take into
account the curved space - time geometry created by the particles themselves.
In other words, gravitational interactions are at least as important as the
rest, and can not be neglected anymore as it is usually the case in particle
physics. In this context, gravitational shock wave backgrounds play a relevant
role as they are the metrics generated by ultrarelativistic particles. The
Aichelburg - Sexl (A - S) geometry\refto{AS71}, is a typical example, as the
ultrarelativistic limit of the Schwarzschild metric. The A - S metric was
generalized to include the electromagnetic momentum of the particle,
by obtaining the ultrarelativistic limit of the Reissner -
Nordstr\"om solution\refto{LS90}. In the paper\cite{LS92a} we find the
ultrarelativistic limit of the Kerr - Newman geometry, to include the charge
and spin of the particles.

In the elementary particle context in which we are studying these metrics,
several interesting features are to be mentioned: Remarkable enough, the
Klein - Gordon equation\refto{H87,dVS89b,LS90} and the string equations
\refto{dVS89a,AK88}
 have been exactly solved in these geometries. The scattering matrix in the A -
S geometry exhibit poles at $iGs=n=0,1,2,...$, where $s$ is the particle
energy\refto{H87}. Such pole structure is also present in the two body
scattering amplitude for the scalar ground state (tachyon) computed from
strings\refto{CdV91}. It must be noticed that the presence of such poles in the
$S$ - matrix is associated to the point - like (structureless) character
of the shock wave source. The contribution of small (zero) impact parameter
yields in this case such poles. Recently, the present authors found the
shock wave spacetimes generated by ultrarelativistic extended sources, as
ultrarelativistic cosmic strings (with charge and spin), and other
topological
defects (as monopoles and domain walls). The scattering matrices of Klein -
Gordon fields in this extended source geometries do not exhibit the
Coulombian type poles
characteristic of the point - like sources, (although, in the cosmic string
 case, the small $s$ and small $t$ behavior give the usual one - graviton
exchange amplitude). In the case of the effective shock wave generated by
Planckian energy string collisions,\refto{AK88} the source is an extended
one\refto{V88,LS92b}, and the scattering matrix does not exhibit
such poles neither.

Another feature in this context is that, for both, point - like as well as
extended sources, the exact $S$ - matrix of the relativistic geometry coincides
with the $S$ - matrix of the weak field limit geometry of
 the static sources, in the large impact parameter regime. In other words, the
$S$ - matrices produced by the ultrarelativistic limit and the weak field
limit of a given geometry coincide.

In the ultrarelativistic limit, the sources travel at the speed of light
(i.e. $\gamma=(1-v^2)^{-1/2}\to 0$) and are massless. In order to describe
the non trivial behavior of the boosted gravitational field in this limit,
the different parameters of interest (mass, electric, magnetic or color
charges, spin) must be chosen to be $\gamma$ - dependent and vanish in a
specific way. The mass goes to zero like $m=p\gamma^{-1}$, the charge
like $Q=p_e\gamma^{-1/2}$ and the spin like $S=pa\gamma^{-1}$, keeping
their respective momenta $p, p_e, a$ fixed. It is the total spin (and not the
spin per unit mass $a$) which becomes $\gamma$ - dependent. The spin vanishes
like the mass, whereas the charge vanishes slower.

Since an ultrarelativistic particle has only two possible spin polarizations,
$\pm S$, in the direction of motion, we apply the Lorentz boost parallel to
the spin. In order to apply a Lorentz boost, it is convenient to start with
the metric expressed in Boyer - Lindquist type coordinates and then to take
asymptotically Cartesian coordinates (as in eq (2) here). The limit
$v\to c$ is subtle, although it is well defined in the sense of distributions
functions. The method to compute this limit consists essentially of
integrating over the null variable $u=z-t$ (the shock wave will be located
at $u=0$), then to take the limit $v\to c$ and finally derivate with respect
to $u$.  Thus, we obtain for the metric profile function
$f(\rho)$ (see eq (9)),
where we observe that the spin $a^2$ contribution couples to the
kinematic momentum $p$ only through even powers of $1/\rho$, whereas the
coupling to the electromagnetic momentum $p_e$ is through odd powers
of $1/\rho$.

In the limit $v\to1$, the terms corresponding to the crossed component
$g_{t\varphi}$ go to zero as $\gamma^{-1}\delta(u)$. As can be seen from
its multipole decomposition, boosting $g_{t\varphi}$ only adds a factor
$\gamma$ due to the boost in $t$ ($\varphi$ coordinates are not boosted).
Thus, the limit will yield terms proportional to $\gamma^{-1}\delta(u)$.

Alternatively, metrics of this type can be found\refto{DH85}
 by starting from flat space - time and
 generating a shock wave through an appropriate shift in the null
coordinate $v=z+t$. In our problem here, since an explicit expression for
the source of the Kerr geometry is not known, it is more convenient to take
the problem all the way around. Thus, from the metric profile function
$f(\rho)$ found here, we are able to obtain the source of the
ultrarelativistic Kerr geometry (see eqs (14) - (15)). These results complete
 and enlarge our previous results presented in \Ref{LS89b}.

We have also studied the effect of the spin in the scattering of test particles
and study the scattering of a Dirac field of spin 1/2 in this shock wave
geometries.\refto{LS92a} The
Dirac equation in this background can be diagonalized to have four uncoupled
components of the spinor, each component fulfilling an equation similar to
that of the scalar field. The $S$ - matrix depends no only on the profile
 function $f(\rho)$, but also on its derivatives with respect to the
 transversal coordinates, due to the dependence on particle's spin.
The poles appearing in the scattering amplitude of a scalar field
in the Aichelburg - Sexl geometry survive, in some way, for the Dirac fields
now considered. In the case of the ultrarelativistic Reissner - Nordstr\"om
metric, the number of such poles is quadruplicated\refto{LS90} and we expect
that for the ultrarelativistic Kerr - Newman geometry this number will
be still larger. Finally, we compared to and point out some features of the
propagation and scattering of fundamental strings.

\head{2. The Ultrarelativistic Kerr - Newman Geometry}
\bigskip

The Kerr - Newman metric in Boyer - Lindquist coordinates reads\refto{2}
$$ds^2=-(\Delta-a^2\sin^2\theta)q^{-2}dt^2-2Mra\sin^2\theta q^{-2}dtd\varphi+
{q^2\over\Delta}dr^2+q^2d\theta^2+$$
$$+\left[(r^2+a^2)^2-\Delta a^2\sin^2\theta
\right]q^{-2}\sin^2\theta d\varphi^2\eqno(1.1)$$
where
$$\Delta=r^2-2Mr+a^2+Q^2\quad,\quad q^2=r^2+a^2\cos^2\theta~.$$
Having in mind to apply a Lorentz boost we write metric \(1.1) in
asymptotically Cartesian coordinates
$$x=(r^2+a^2)^{1/2}\sin\theta\cos\varphi~,$$
$$y=(r^2+a^2)^{1/2}\sin\theta\sin\varphi~,\eqno(1.2)$$
$$z=r\cos\theta~~~,~~~t=t~.$$
We now apply a Lorentz boost (with velocity $v$)
in the z-direction (an ultrarelativistic particle have only two
spin polarizations ($\pm S$), in the direction of motion)
$$\tilde t=\gamma(t+vz)\quad,\quad\tilde z=\gamma(z+vt)\quad,$$
$$\tilde y=y\quad,\quad\tilde x=x\quad,\eqno(1.3)$$
$$\gamma=(1-v^2)^{-1/2}$$
The limit $v\to1$ in the resulting metric is a delicate point, but we can
extract a meaningful expression from it by letting $M ~,~ Q$ and $S$ be
$\gamma$ - dependent and going to zero as
$$M=p\gamma^{-1}~~,~~Q^2=p_e^2\gamma^{-1}~~,~ ~S=aM=ap\gamma^{-1}~.\eqno(1.4)$$
Thus, after use of eqs\(1.2)-\(1.4) in metric\(1.1) we find
$$ds^2=-d\tilde t^2+d\tilde z^2+d\tilde y^2+d\tilde x^2+
(d\tilde t-vd\tilde z)^2\lim_{v\to1}\left[\gamma^2(\Delta g_{\tilde t\tilde t}+
\Delta g_{\tilde z\tilde z})\right]+O(\gamma^{-1})~,\eqno(1.5)$$
where
$$\Delta g_{\tilde t\tilde t}={2M\tilde r-Q^2\over
\tilde r^2+a^2\cos^2\theta}~,$$
$$\eqno(1.6)$$
$$\Delta g_{\tilde z\tilde z}={(2M\tilde r-Q^2)(\tilde r^2+a^2)\cos^2\theta
\over (\tilde r^2+a^2\cos^2\theta)(\tilde r^2-2M\tilde r+a^2+Q^2)}~,$$
and
$$\cos\theta=z(\tilde z)/\tilde r~,$$
$$\tilde r^2={1\over2}\left[z(\tilde z)^2+\rho^2+\sqrt{(z^2(\tilde z)
+\rho^2)^2+4a^2
z^2(\tilde z)}\right]~,\eqno(1.7)$$
$$\rho^2=\tilde x^2+\tilde y^2-a^2~.$$
Let us notice that the terms corresponding to $g_{t\varphi}$ in the limit
$v\to1$ go to zero as $\gamma^{-1}\delta(\tilde t-\tilde z)$. This is
sketched in the appendix of \Ref{LS92a}.

The limits appearing in eq\(1.5) are computed explicitly in the Appendix
of \Ref{LS92a} giving rise to a shock wave metric
$$ds^2=du~dv-f(\rho)\delta(u)du^2+d\tilde y^2+d\tilde x^2~,\eqno(1.8)$$
where
$$f(\rho)=-8p\ln\rho-{3\over 2}\pi p_e^2\rho^{-1}+p\sum_{m=1}^{\infty}
B_m(a^2/\rho^2)^m-\pi p_e^2\rho^{-1}\sum_{m=1}^{\infty}
C_m(a^2/\rho^2)^m\eqno(1.9)$$
where $u=\tilde z-\tilde t$ and $v=\tilde z+\tilde t$ are the usual null
coordinates and the coefficients $B_m$ and $C_m$ are numerical factors
which are found in the Appendix of \Ref{LS92a}
$$B_m=4\sum_{k=0}^m(-)^{m-k}\sum_{j=0}^{m-1}(-)^j\pmatrix{m-1\cr j\cr}
\left\{{A_{k,2m-2k+1/2}\over 2j+2m+1}+{A_{k,2m-2k+3/2}\over 2j+2m+2}\right\}~,
\eqno(1.10)$$

$$\eqalign{C_m=&\sum_{k=0}^m(-)^{m-k}4^{-m}
\bigg\{{A_{k,2m-2k+1}\over(2m)!}(4m-4)!!
\prod_{j=0}^{m-1}{2m-2j-1\over2m+2j+1}+\cr
&+{A_{k,2m-2k+2}\over2(2m+1)!}(4m+1)!!
\prod_{j=0}^{m}{2m-2j+1\over2m+2j+1}\bigg\}~,\cr}
\eqno(1.11)$$
The metric\(1.5) represents the gravitational field of a boosted particle
of mass $M$, charge $Q$ and spin $S$ in the ultrarelativistic
 limit $v\to1$ when all these
three quantities go to zero keeping their respective associated momenta
($p,~p_e~,~{\rm and}~a$) fixed. The $\rho$ dependence of each contribution is
different. The spin contribution couples to the kinetic momentum only through
even negative powers of $\rho$ whereas the coupling to the charge is through
odd negative powers of $\rho$.

The spacetime is flat everywhere except on the null plane orthogonal to the
 direction of propagation and has a $\delta(u)$-behaviour characteristic of
such light-like contractions.

In the metric \(1.8)-\(1.9), the condition $\rho^2=\tilde x^2+\tilde y^2-a^2
\geq0$ represents that this boosted solution ``remembers'' the ring singularity
of the original Kerr metric.

Let us notice that the profile function \(1.9) remains invariant after a change
in the orientation of the spin, i.e. $a\to -a$. This is due to the fact that
in the ultrarelativistic limit $g_{t\varphi}\to0$, which is the only spin -
direction dependent term.

We should also point out that the boosted geometry  \(1.8)-\(1.9), has
$M^2=p^2\gamma^{-2}<Q^2+a^2=p_e^2\gamma^{-1}+a^2$ when $\gamma\to\infty$. This
means that metric\(1.1) represents not a black hole, but a point-like particle
at the irremovable singularity $r=0$. This is in agreement with the
interpretation and the elementary particle context in which we are using
this geometry: the ratio $(Q^2+a^2)/M^2$ is always bigger than one (This is
not the case, of course, for astrophysical objects). The solutions
representing black holes with $M^2>Q^2+a^2$ do not exist in the
ultrarelativistic limit.

It is worth to remark that recently\refto{10},  only the first term of eq\(1.9)
was obtained for
the ultrarelativistic limit of the Kerr
metric. The authors of \refto{10}
do not reach to find the additional powers of $\rho$ proportional to $B_m$,
because they set $u=0$ before taking the singular limit $v\to1$. If we had
made this in eqs (47) and (53) all the additional terms would had gone to
 zero,
but in this problem one has to take a double limit,
$v\to1$ and $u\to0$, which must be
treated in a distributional sense, that is, as we have done here,
in the Appendix of \Ref{LS92a}. The final
 result will indeed
be that given by eqs\(1.9)-\(1.11).

\head{3. The Ultrarelativistic Energy - Momentum Tensor}

The metric\(1.8) can also be obtained from flat spacetime by generating a
 gravitational
shock wave through a shift in the null coordinate $v$, following on the
lines of the procedure developed
by Dray and 't Hooft\refto{DH85}. This procedure consist of solving the
Einstein
equations for a null source representing a particle travelling at the speed
of light. This procedure can be generalized
to D - dimensional curved backgrounds including sources and
cosmological constant\refto{LS89a,ks94}.

This formalism will allow us to study the gravitational shock waves in a
curved background:
$$ds^2=2A(u,v')dudv'+g(u,v')h_{ij}(x^k)dx^idx^j~;\eqno(iii.1)$$
solution of Einstein equations without shock wave.
$$R_{\mu\nu}-{1\over2}g_{\mu\nu}R+\Lambda g_{\mu\nu}=T_{\mu\nu}\eqno(iii.2)$$

Then, let us make the following ans\"atze for the metric solution of Einstein
eqs., including the shock wave as source:
$$ds^2=2A(u,v)du[dv-f(x^k)\delta(u)du]+g(u,v)h_{ij}(x^k)dx^idx^j\eqno(iii.3)$$

The source is given by
$$T_{\mu\nu}=T_{\mu\nu}^{(b)}+T_{\mu\nu}^{(p)}~,$$
where
$$T_{uu}^{(p)}=\sigma(x^k)\delta(u)~,\eqno(iii.11)$$
The proposed metric \(iii.3) will be a solution of Einstein equations if:
$$\partial_vA(0,v)=0=\partial_vg(0,v)=0=T_{vv}(0,v)~,\eqno(iii.17)$$
$${A\over g}\nabla^2_{D-2}f-(D-2){g,_{uv}\over2g}f=\sigma(x^k)\eqno(iii.18)$$
both evaluated at $u=0$.
Examples of application in a flat background, i.e.
$A={1\over2} {\rm ~~~and~~~} g=1~,$ can be computed for an uncharged particle
with energy-momentum tensor given by
$T_{uu}=p\delta(\rho)\delta(u)~.$
Plugging this expression into\(iii.18):
$$\nabla^2_\rho f=16\pi p\delta(\rho)~.\eqno(iii.20)$$
we find the solution to be
$$f(\rho)=8p\ln\rho~.\eqno(iii.21)$$
i.e., the Aichelburg - Sexl metric. The cases of a charged and spinning
particle is discussed in Refs. \cite{LS90} and \cite{LS92a}.

This method allows to obtain the metric function $f(\rho)$
given the $T_{\mu\nu}$ corresponding to the ultrarelativistic particle;
in the Kerr-Newman case, however,
it is more convenient to take the problem all the way around since an analytic
expression for the exact source
of the Kerr black hole remains unknown. Thus, given $f(\rho)$ by the expression
\(1.9) we can find  $T_{\mu\nu}$, which in our case has only one component
different from zero
$$T_{uu}=\delta(u)\nabla^2_\perp f(\rho)=\delta(u)\rho^{-1}\partial_\rho(
\rho\partial_\rho f)\eqno(3.3)$$
Plugging \(1.9) into this equation we obtain
$$\eqalign{T_{uu}=\delta(u)\bigg\{p&\delta(\rho)+4p\rho^{-2}
\sum_{m=1}^{\infty}m^2B_m(a^2/\rho^2)^m+\cr
+&{3\over 2}\pi p_e^2\rho^{-3}+\pi p_e^2\rho^{-3}
\sum_{m=1}^{\infty}(2m+1)^2
C_m(a^2/\rho^2)^m\bigg\}\cr}\eqno(3.4)$$

Here we can identify the first addend as the source of an ultrarelativistic
 particle with
momentum p. The third addend, as we have shown\refto{LS90}, corresponds to the
contribution of the electromagnetic field generated by a point charge boosted
to
the speed of light. The other two contributions are the new terms which bring
the effect of rotation. Clearly, we can distinguish between them: one is
the effect of the Kerr metric (proportional to the kinetic momentum $p$)
 and the other is the effect of
 the electromagnetic field in rotation (proportional to $p_e$).

Let us observe that the term coming from the Kerr metric can be written as
$$T_{uu}^{Kerr}=\sigma(\rho)\delta(u)\eqno(3.5)$$
Which allows to be interpreted as a ``shell'' of null fluid moving in the
$v$ direction with an axially symmetric surface energy density given by
$$\sigma(\rho)=p\left\{\delta(\rho)+4\rho^{-2}\sum_{m=1}^{\infty}m^2
B_m(a^2/\rho^2)^m\right\}\eqno(3.5')$$
Of course, this expression for $T_{uu}$ is the Kerr's source already boosted
to the speed of light. We can not find uniquely from it the non-boosted
$T_{\mu\nu}^{Kerr}$ (subject which would deserve the effort), because some
information was lost in the process of taking the limit $v\to c$.
Remarkably enough Balasin and Nachbagauer\refto{BN93} in a series of papers
have obtained the energy momentum tensor-distributions by using the Kerr-
Schild decomposition of the metric successively
for the  Schwarzschild, Kerr and Kerr-Newman geometries.

\head{4. The Metric of Ultrarelativistic Particles and Extended Sources}
\bigskip

As we have seen, the metric created by an ultrarelativistic source can
be described by
$$ds^2=dudv+f(x^i)\delta (u)du^2+dx^idx^j~~~,~~~i,j=2,3,.....D-1\eqno (b1.1)$$
where $x^i$ are the coordinates transversal to the motion which takes
place in the plane $u,v$.
$$u= x^1 - t\quad , \quad v= x^1 + t$$
This represents a gravitational shock wave located at the null plane $u=0$.
Outside the plane $u=0$ this metric is flat. It is completely characterized
by the function $f(x_{\bot})$. For an uncharged and spinless point
particle of momentum $p$ in four dimensions
$$f(x_{\bot})=8Gp\ln{\rho}~~~~,~~~~\rho ^2=\sum_{i=2}^3x^ix^i\eqno(2.1)$$
This is the Aichelburg - Sexl (A - S) metric\refto{AS71}. This metric
 can be obtained by applying a Lorentz boost to the Schwarzschild
geometry with the mass going to zero as
$M=p\gamma^{-1}~~,~~\gamma^{-2}=(1-v^2)$
and finally, taking the velocity  tending to
the speed of light $(v\to 1)$.
  By the same method it can be obtained the form of the function $f$ when
one starts from a particle originally endowed of charge\refto{LS90} (Reissner -
Nordstr\"om metric) and spin\refto{LS89b,LS92a} (Kerr - Newman metric)
$$f(\rho)=-8Gp\ln{\rho}-{3\over 2}\pi p_b^2{1\over \rho}+p\sum_{n=1}^{\infty}
B_n \left( {a^2\over \rho^2}\right) ^n+{\pi p_b^2\over \rho}\sum_{n=1}^{\infty}
C_n\left( {a^2\over \rho^2}\right) ^n\eqno(2.2)$$
where $p_b$ is the impulse due to the electromagnetic field and $a$ is the
intrinsic angular momentum per unit mass associated to the particle. $B_n$ and
$C_n$ are determined numerical constants.

There is an equivalent method\refto{DH85,LS89a} to obtain $f(\rho)$. This
consists of writing down the Einstein equations for a metric like \(b1.1). This
ansatz reduces the Einstein equations to only one relevant
linear equation
$$\delta(u).\bigtriangledown_{x_{\bot}}^2f(x_{\bot})=2T_{uu}\eqno(2.3)$$
$T_{uu}$ being the only non zero component of the matter source in the
ultrarelativistic limit. For an infinite thin shell of matter propagating at
the speed of light: $T_{uu}=\sigma (x_{\bot})\delta (u)$ ,
 $\sigma (x_{\bot})$ being the
matter density of the extended source.

Let us suppose now that instead of a point - like {\it particle}
we proceed to boost a
{\it cosmic string}. What we would obtain as result? The answer is\refto{LS91}
 again a metric like \(b1.1), but now $f(x_{\bot})=f(y)$.
 i.e. the function $f$ will
only depend on the distance $y$ perpendicular to the string
instead of the radial
distance to the particle $\rho$, which is reasonable, if we take into
 account the symmetries of the problem.
By replacing in eq \(2.3) the expression for the $T_{\mu\nu}$ corresponding
to cosmic strings, we find\refto{LS91}
$$f(y)=2p_{l}\mid y\mid\quad,\eqno(2.4)$$
for {\it gauge} or {\it local} cosmic strings laying along the {\bf z} axis.
 We have used
that $T_t^t=T_z^z=\mu\delta(x)\delta(y)$, and when boosted in the {\bf x}
direction with $v\to 1$ this $T_{\mu\nu}$ reduces to
$$T_{uu}=p_l\delta (y)\delta(u)\quad.\eqno(2.5)$$

It is worth to remark that for a local string in a $D$ - dimensional
spacetime, we had obtained:
$$f(y)={4Gq\Gamma[(D-5)/2]\over(2\pi)^{(D-5)/2}}\mid\vec y_{D-3}\mid^{5-D}~,
\eqno(D)$$
where $\vec y_{D-3}$ is the $D-3$ dimensional vector perpendicular to the
string and to the direction of motion. Eq\(D) gives the same result than
for the case of an ultrarelativistic particle in $D-1$ dimensional
spacetime (Aichelburg - Sexl metric, see eq(26)).

The case of {\it global cosmic strings} with
energy momentum tensor given by
$$T_t^t=T_z^z=T_r^r=T_{\theta}^{\theta}=-{\mu\over 2r^2}\eqno(2.6)$$
when boosted at $v\to 1$ gives
$$T_{uu}=-p_g\mid y\mid ^{-1}\delta (u)~~~,~~~y\not= 0~.\eqno(2.7)$$
This, from eq \(2.3) yields
$$f(y)=p_g\mid y\mid (ln\mid y\mid -1)~~~,~~~y\not= 0 .\eqno(2.8)$$

For the purposes of further discussion, it is interesting to see the metric
generated by {\it Domain Walls}.
Their formation is associated to a discrete symmetry breaking. For a wall
located in the $ x$ - $y$ plane, the energy momentum tensor can be represented
by\refto{LS91}
$$T^0_0 = T^x_x = T^y_y = \sigma\delta (z)\quad\quad ,\quad\quad T^z_z = 0
\eqno(III.14)$$
where $\sigma$ is the energy scale of symmetry breaking corresponding to
domain walls.

After boosting in the {\bf z} - direction and by
taking
$$\sigma = p_{\sigma}\gamma ^{-1}\quad ,\eqno(III.15)$$
in the ultrarelativistic limit we have
$$T_{uu} = p_{\sigma}\delta (u)\quad\quad ,\quad\quad u = z - t\quad
,\eqno(III.16)$$
and all other components vanishing. The metric function is given by

$$f(\rho )={1\over 2}p_{\sigma}\rho ^2\eqno(gua)$$

\head{5. The Scattering of Quantum Fields by these Gravitational Shock Waves}

Let us study now the Klein-Gordon  equation in the ultrarelativistic
Reissner - Nordstr\"om geometry discussed before. This describes the following
physical process: the scattering of two spinless particles (1 and 2) of
masses $m_1~,~m_2=m$; both being $\ll m_{pl}$, and charges $e_1=0$,
$e_2=e$, at energies of the order or larger than Planck mass $m_{pl}$.

Following \ref{H87} we choose a frame where only one of the particles
(particle 2) has an energy of the order of $m_{pl}$. Therefore the particle 2
is ultrarelativistic momenta $p$ and $p_e$ respectevely and the space-time
around it is described by generalized AS geometry discussed above.
(The motion of particle 2 is taken here along the $x$ axis with speed $+1$).
Particle 1 is a test particle since its energy is much smaller than $m_{pl}$.
Its dynamics is described by the Klein-Gordon equation in the geometry of
Eq.\(b1.1).

By writing the wave function as
$$\Psi(u,v,x^k)=\exp(-i\omega v/4)\Phi(u,x^k)~,\eqno(iv.3)$$
we have (since $v$ is a cyclic variable)
$$\partial_u\Phi=\left[-i\omega/4 f(x^k)\delta(u)+i/\omega(
\nabla_{\perp}^2-m^2)\right]\Phi~.\eqno(iv.4)$$
Let us take an incoming plane wave for $u<0$.
Thus, for $u>0$, the continuous regularization produces the following
outcoming wave function\refto{dVS90}
$$\Phi_>(u,x^k)=\int dp_{\perp}^{D-2}\exp\left\{i\vec p_{\perp}\cdot
\vec x_{\perp}-iu/\omega(p_{\perp}^2+m^2)\right\}S(k_{\perp},p_{\perp},
\omega)\eqno(iv.6)$$
where
$$S(k_{\perp},p_{\perp},\omega)=(2\pi)^{2-D}\int dx_{\perp}^{D-2}\exp
\left[i(\vec k_{\perp}-\vec p_{\perp})\cdot
\vec x_{\perp}+i\varphi_D(x^k)\right]~,\eqno(iv.7)$$
$S(k_{\perp},p_{\perp},\omega)$ is the $S$-matrix which transforms an incoming
state with momentum $(k_{\perp},\omega)$ into an outgoing state of momentum
$(p_{\perp},\omega)$. $\varphi_D(x^k)$ is the phase shift suffered by the
incoming wave, that happens to be the only relevant effect since there is not
particle creation.
$$\varphi_D(x^k)={\omega\over4}f(x^k)~.$$

As examples of application of these results we can compute the S-matrix
of a profile function $f(\rho)$ corresponding to the Aichelburg - Sexl metric.
In this case we obtain\refto{H87}

$$S^{AS}(s,t)={1\over\pi}\left(4\over\mid t\mid\right)^{iG(s-m^2)}{\Gamma[1
-iG(s-m^2)]\over\Gamma[iG(s-m^2)]}+\delta(\vec k_{\perp}-\vec p_{\perp})~.
\eqno(iv.11)$$
where
     $$Gs=\omega p+Gm^2~~,~~t=-\mid\vec k_{\perp}-\vec p_{\perp}\mid~.$$
It can be seen that there is no particle production\refto{GV91}(the Bogoliubov
coefficients $\beta_{kp}$=0).

$S^{cont}$ exhibits an infinite sequence of imaginary poles in
$$G(s-m^2)=-in~,~n=1,2,3,..\eqno(pt)$$
We can observe a similarity between eq \(iv.11) and the Veneziano amplitude
for strings.

In the same way, we can compute the S-matrix for the profile corresponding
to the Reissner - Nordstr\"om metric\refto{LS90}.

This case contains the later and, in fact, it can be written as a superposition
$$S(s,t)=\sum_{n=0}^{\infty}{1\over n!}\left[{iG(s-m^2)}\over 4p
\right]^nS^{\alpha=0}\left[iG(s-m^2)+n/2,t\right]~.\eqno(iv.14)$$
This series exhibits an infinite sequence of imaginary poles in
$$iG(s-m^2)=in/2~~,~~n {\rm ~integer}~.\eqno(iv.15)$$
We thus observe the very interesting feature that the electromagnetic momentum
$p_e$, increases the number of poles by a factor four. From the residues of
these poles, we see that for $p_e=0$, the possible angular momenta for the
bound state ``resonances" are restricted by $n$. For $p_e\neq0$, whatever the
resonance be, all angular momenta are included (all spin are allowed).

\head{6. Background Metric for the Scattering of Superstrings. Large and
Intermediate Impact Parameters}

Amati, Ciafaloni and Veneziano\refto{ACV88} have obtained an expression
for the scattering amplitude of superstring collisions in flat space time,
valid for high energies (i.e. energies much larger that ${1/\sqrt{\alpha '}}$).
The high energy regime investigated is that of large effective coupling
$g^2\alpha ' s/\sqrt{\alpha '}^{D-4}$ and small loop expansion parameter
$g^2/\sqrt{\alpha '}^{D-4}~,$ $(\alpha 'g^2=16\pi G)~.$
By resumming in the eikonal approximation, the leading behaviour
 of the string loop diagrams
in not compactified D-dimensional flat spacetime,
 they found an S-matrix:
$$S=\exp\left( 2i\int_0^{\pi}:a(s,b+\hat x^u(\sigma _u,0)-\hat x^d(
\sigma _d,0)):{d\sigma _ud\sigma _d\over \pi^2}\right)\eqno (b1.3)$$
where $b$ is the impact parameter between the center of mass of the two
colliding strings. $x^u$ and $x^d$ are the oscillations around the center
of mass of the string at rest; $s$ is the usual Mandelstam variable
and $a$ is the string tree amplitude
$$\eqalign{a(s,b)=&{Gs\over 2\pi ^{D/2-2}}b^{4-D}\int_0^{g(b,s)}
dte^{-t}t^{D/2-3}\cr
g(b,s)=&{b^2\over 4\alpha '
[ln(s)-i\pi /2]}={b^2\over\Delta^2}~,\cr}\eqno (b1.4)$$
where the above integral can be re-expressed in terms of the
incomplete gamma function
$\Gamma\left[D/2-2,g(b,s)\right]$ and $\alpha ' $ is the usual string
slope.

On the other hand, the scattering of free strings\refto{AK88,dVS89b}
in the shock wave backgrounds \(b1.1)
 is exactly solvable. For a free superstring in this background
 the S-matrix has the expression\refto{dVS89b}
$$S=\exp{\left( {i\over 2\pi}p^u\int_0^{\pi}d\sigma f\left( x(\sigma ,0)
\right)\right)}\eqno (b1.2)$$
where $p^u$ is the impulse associated to the string, $\sigma$ is the
world sheet spatial coordinate and $x(\sigma,\tau=0)$ is the projection
of the D-dimensional superstring coordinate on the plane $u=p^u\tau=0$ where
the shock wave is located
(perpendicular to the motion of the shock wave ultrarelativistic source).

 Now, if we want both S-matrices describe the same phenomena, we
equate expressions \(b1.2) and \(b1.3). One obtains in this way the form
of the function $f$ appearing in the metric\(b1.1)
$$f(\rho)={8q\over s}\int_0^{\pi}:a(s,\rho -\hat
x^d(\sigma _d,0)):{d\sigma_d
\over\pi}\eqno(b1.5)$$
where $q$ is the impulse associated to the shock wave and $s=2p^uq$.

Plugging eq \(b1.4) into \(b1.5) we obtain for $f(\rho)$:

$$f(\rho )=C\int _0^{\pi }{d\sigma\over\pi }:b^{4-D}\int_0^{g(\rho)}dte^{-t}t^
{D-6\over 2}:\eqno (3.1)$$

where
 $$g(\rho)={b^2\over 4\alpha '(ln(s)-i\pi /2)}~~,~~
C={4Gq\over 2\pi^{D/2-2}}~~,~~
b=\rho-x^d(\sigma ,0)\eqno (3.2)$$

In the regime
studied here, $b$ never goes to zero ($b\gg x^d\sim$ string size).
In the regime of large impact parameters described by
$$x^d\ll\sqrt{\alpha '\ln s}\ll b\quad,\eqno(1)$$
where $s$ is the usual  Mandelstam variable and $x^d$  the characteristic size
of string fluctuations, the effective geometry generated by flat space string
collisions is the Aichelburg - Sexl (A - S) geometry. This is the gravitational
shock wave produced by an ultrarelativistic uncharged and spinless point
particle in $D$ dimensions.

We investigate here the regime
$$x^d\ll b\ll\sqrt{\alpha '\ln s}\eqno(2)$$

In this regime (let us call  it small-intermediate $b$), $b$ is still  much
bigger than the characteristic size of string excitations, so that strings
do not superpose each other. We find that it is possible to associate an
 effective curved metric to the string collision process in this regime. This
is a gravitational shock wave, but with a profile function $f$ given in
$D=4$ by
$$f(\rho)\simeq{C\over(D-2)\Delta^{D-2}}\rho^2\quad,\quad\Delta^2=4\alpha'
(\ln(s)-i\pi/2)~~,~~
C={4Gq\over 2\pi^{D/2-2}}~~{\rm (see~below)}~.\eqno(3)$$
This metric is generated by an extended source with energy density distribution
$$\sigma (\rho )={2C\over\Delta^{D-2}}\exp\{-\rho^2/\Delta^2\}~,\eqno(ss)$$
$q$ being the kinematic momentum.
This dependence on $\rho$ in $f$ and $\sigma$ is characteristic of the string
collision process at intermediate $b$, and reproduces at first order
other simpler
extended sources as domain walls\refto{LS91} or "homogeneous
beams"\refto{17,V88}.
Notice also the difference with the point-like particle A - S geometry, for
which $f(\rho)=8Gp\ln\rho$ (in four dimensions) and $\sigma(\rho)=8Gp\delta
(\rho)$, ($p$ being the momentum of the particle). The scattering collision
that generates the shock wave no longer behaves as an effective point-like
 source in the scattering at the small - intermediate $b$.

We study the scattering of test particles and fields by this geometry. The
 deflection angle is (in D=4)
$$\Theta=\cases{8Gp/b &for~large $b$\cr
4Gqb/\Delta^2 &for~intermediate $b$.\cr}\eqno(6)$$

The $S$ - matrix of Klein - Gordon fields in this geometry is analysed in
order to describe some features of the ultra - high energy string - string
scattering as one goes along decreasing impact parameters $b$. The whole
picture is depicted in the Figure 1 of ref. \cite{LS92b}.
In this scattering matrix, the poles
$iGs=n+1~,~ n=0,1,2,...$ do not appear. Such poles, appearing in the
$S$ - matrix of
a particle (or string) in the A - S geometry, arise because of the extrapolated
use of the A - S geometry down to small impact parameters $0\leq b\leq x^d$
(including $b=0$). For intermediate $b$, deviations from the A - S geometry
are important and, on the other hand, for very small $b$ ($0\leq b\leq x^d$),
the use of the external metric approximation is not valid.

Let us study this function $f(\rho)$ in the case where
$$\rho\gg [\alpha 'ln(s)]^{1/2}\gg x^d\quad .\eqno (3.2')$$
This case corresponds to the scattering of superstrings at very high energies
, but with large impact parameter. As $\rho$ is large, the integral in the
 variable t in eq \(3.1), can be taken between zero and infinity (due to the
exponential in the integrand). Also, as $\rho\gg x^d(\sigma)$ we can take
$b^{4-D}$ out the integral. Then,
$$f(\rho )=C\Gamma\left(D/2-2\right)\rho ^{4-D}\quad,\eqno (b3.3)$$
which corresponds to
the A - S metric (uncharged and spinless particle) in D dimensions.

The other limit we want to study is
$$[\alpha 'ln(s)]^{1/2}
\gg \rho\gg x^d(\sigma)\quad ,\eqno(3.3')$$
which in
some sense represents the opposite to the last limit studied, because it
 corresponds to the scattering of strings at very high energies and at smaller
(intermediate, say)
 impact parameters (although not so small to superpose the superstrings each
other). The regime given by eq \(3.3') means that the impact parameters
 are
smaller than those considered in the regime of eq \(3.2'), but still bigger
than the characteristic size of fundamental string excitations.

Here, it is convenient to analyse
the effective or equivalent energy - momentum tensor, $T_{uu}$, instead of
the metric function $f(\rho)$. To do this we use eq \(2.3)
$$T_{uu}(\rho )={1\over 2}\delta (u){1\over \rho ^{D-3}}\partial _{\rho }
\left (\rho ^
{D-3}\partial _{\rho }f(\rho )\right )={2C\delta (u)\exp\{-\rho^2/\Delta^2\}
\over\Delta^{D-2}}=R_{uu}\eqno (b3.4)$$

This expression is exact within the approximations posed to obtain \(3.1).
Here, can be observed clearly the meaning the
regimes \(3.2') and \(3.3') that we
have considered. In the limit \(3.2'), $T_{uu}\to0$, which characterizes
the Aichelburg - Sexl solution (represented by a $\delta(\rho)$). While in
the limit \(3.3'), (for $s$ fixed)
$T_{uu}\sim$ constant (Let us observe that this is the case of domain walls).

In terms of a metric to which we apply a Lorentz boost with $v\to 1$,
$T_{uu}\sim$ constant
would represent a particle that develops a non - localized interaction
 term.
To this $T_{\mu\nu}$ corresponds a metric function
$$f(\rho)\simeq{C\over(D-2)\Delta^{D-2}}\rho^2~~,~~
C={4Gq\over 2\pi^{D/2-2}}~~,\quad\rho<\Delta ,
\eqno(3.6')$$
 which differs from
the Aichelburg - Sexl metric (see eq\(b3.3)).
 This new dependence arises as a consequence of the non-point-like interaction
of strings in the small - intermediate impact parameter regime.
 One expects that the superstring collision that
generates the  effective shock wave geometry
no longer behaves as an effective point like source when
 studying the scattering at smaller impact parameters.
It is worth to remark that the energy density \(b3.4) is proper of the
scattering of fundamental strings. It contains the homogeneous beam
 approximation\refto{17,V88}
$$\sigma_h=\cases{\sigma_0 &if $\rho<\Delta$\cr
0 &if $\rho>\Delta$.\cr}$$
It is also worth to point out that the presence of this energy density is
enough to remove the so called 't Hooft poles\refto{H87,H88}. This displays
the ``softener" character of fundamental strings, already noted in other
 computations within the theory of fundamental strings.

\head{7. On the Universality of 't Hooft's Poles}

Let us consider now the $S$ - matrix of Klein - Gordon fields in the spacetime
geometry of eq\(b1.1) in four dimensions. The $S$ - matrix is given
by\refto{dVS89a}
$$S({\bf k}_{\perp},{\bf p}_{\perp},\omega)=\int{{d\rho\over 2\pi }\rho
J_0(\mid{\bf k}_{\perp}-{\bf p}_{\perp}\mid\rho)e^{{-i\omega\over 4}f(\rho )
}}\quad ,\eqno(4.21)$$
In order to use this $S$ - matrix to describe some of the features of the
ultra high - energy string collisions, the integral in this context must be
understood as the sum of at least three integrals
$$S= S_I + S_{II} + S_{III} = \int _0^{\rho _1} + \int _{\rho _1}^{\rho _2}
+ \int _{\rho _2}^{\infty}\eqno(4.22)$$
In the last integral (large impact parameter regime), we can use the A - S
metric, i.e. $f_{III}(\rho)=8Gp\ln(\rho)$. In the second integral (intermediate
region), $f_{II}(\rho)\simeq{2Gq\over\Delta^2}\rho^2$
 must be plugged in. Unfortunately, it
seems not easy to find a closed expression for the second integral, while for
 the first region the use of the external effective  metric approximation
is not valid (the collision is so close that strings "touch" each other). The
whole picture is depicted in Figure 1 of ref. \cite{LS92b}.
The poles $iGs=n+1~,~n=0,1,2,...$
appearing in the $S$ - matrix of a particle (or string) in the A - S geometry
arise because one usually extends the third integral down to impact parameters
$\rho _2 =0$.

 This integral is given by
$$S_{III}(s,t,\rho _2)={1\over 2\pi}\int _{\rho _2}^\infty d\rho \rho ^{1-2iGs}
J_0(\rho\sqrt{-t})\eqno(5.1)$$
where $t=-\mid{\bf k}_{\perp}-{\bf p}_{\perp}\mid ^2\quad,\quad Gs=\omega p$
and we have taken $f_{III}(\rho)=8Gp\ln\rho$. Then, we find
$$S_{III}={1\over 2\pi}(-t)^{iGs-1}\{-2iGs\rho _2J_0(\rho _2)S_{-2iGs,-1}
(\rho _2)+\rho _2J_1(\rho _2)S_{1-2iGs,0}(\rho _2)\}\eqno(5.2)$$
where $J_0$ and $J_1$ are Bessel functions and
$$S_{\alpha ,\beta}(z)=s_{\alpha ,\beta}(z)+\eqno(5.3)$$
$$+2^{\alpha -1}
\Gamma({\alpha -\beta +1\over 2})\Gamma({\alpha +\beta +1\over 2})
\left[\sin({\alpha
-\beta\over 2}\pi)J_\alpha (z)-\cos({\alpha -\beta +1\over 2}\pi)
N_\alpha (z)\right]$$
\noindent
$S_{\alpha ,\beta}(z)$ and $s_{\alpha ,\beta}(z)$ are Lommel's functions\refto
{19} and can be related to hypergeometric series
$$s_{\alpha ,\beta}(z)={z^{\alpha +1}\over (\alpha +\beta +1)(\alpha -\beta
+1)}
\quad _1F_2\left(1;{1\over 2}(\alpha -\beta +3);{1\over 2}(\alpha +\beta +3);
-z^2/4\right)\quad,\eqno(5.4)$$
$$N_\alpha (z) = \csc(\alpha\pi)[J_\alpha (z)\cos(\alpha\pi)-J_{-\alpha}(z)]
\quad.$$
For $\rho _2\to 0$ , eq\(5.2) gives the A - S scattering
amplitude\refto{H87,dVS89b}.
For any $\rho _2$ finite and non zero, $S_{III}$ do not have the poles at
$iGs=n+1 ,\quad n=0,1,...$. For the purposes of seeing this it is more
 convenient to make the decomposition:  $\int _{\rho _2}^\infty =
\int _{0}^\infty -\int  _0^{\rho _2}$ in eq\(5.1).
 The first term is given by\refto{H87,dVS89b}
$${1\over 2\pi}2^{1-2iGs}
(-t)^{iGs-1}{\Gamma (1-iGs)\over\Gamma (iGs)}\eqno(5.5)$$
while the second term yields
$${1\over 2\pi}\int _0^{\rho _2}d\rho \rho ^{\mu}
J_0(\rho\sqrt{-t})={1\over 2\pi}(-t)^{-(\mu +1)/2}
\sum _{k=0}^\infty {(-1)^k\over 4^k(k!)^2}
{\rho _2^{\mu +2k+1}\over\mu +2k+1}\quad,\quad\mu=1-2iGs\eqno(5.6)$$
where we have used
$$J_0(z)=\sum _{k=0}^\infty (-1)^k{z^{2k}\over 4^k(k!)^2}\eqno(5.7)$$
Eq\(5.6) have poles at $\mu = -(2k+1)$ , i.e. $iGs=n+1 ,\quad n=0,1,...$
, the same than eq\(5.5). Let us study if both expressions have the same
 residues. By multiplying eq\(5.6) by $\mu +2k+1$ and taking the limit
$\mu\to -(2n+1)$ we obtain:
$$Res\left\{\int _0^{\rho _2}\right\} = {1\over 2\pi}(-t)^{-n}{(-1)^n
\over 4^n(n!)^2}\quad,\eqno(5.9)$$
On the other hand, by using that
$$Res\left\{\Gamma (-n)\right\} = {(-1)^n
\over n!}\quad,\eqno(5.10)$$
we obtain for the residue of eq\(5.5) the same expression eq\(5.9). Thus,
in eq\(5.1) the poles disappear.


\head{8. String Instabilities in Black Hole Spacetimes}

The next step in studying the scattering processes at Planckian scales is to
consider, since the beginning, a theory which is considered well suited for
dealing with such huge energies and tiny lengths. So far, the theory of
strings provides a consistent frame where to study the very high energy
scattering processes. In particular, we will discuss the collision (or better
the infall) of a fundamental string in the gravitational background of a
charged black hole.

The study of the string dynamics in curved space-times, reveals new insights
with respect to string propagation in flat space-time (see for example refs
[\cite{vs87,vs88,sv90,gsv,vs91,vms}]).

The equations of motion and constraints for strings in curved spacetimes
are highly non linear (and, in general, not exactly solvable). In ref
[\cite{vs87}], a method was proposed (the ``strong field expansion") to
study systematically ( and approximately), the string dynamics in the
strong curvature regime. In this method, one starts from an exact particular
solution of the string equations in a given metric and then, one constructs
a perturbative series around this solution. The space of solutions for the
string coordinates is represented as
$$X^A(\sigma,\tau)=q^A(\sigma,\tau)+\eta^A(\sigma,\tau)+
\xi^A(\sigma,\tau)+.....~~,\eqno(i1)$$
$A=0,...,D-1$. Here $q^A(\sigma,\tau)$ is an exact solution of the string
equations and $\eta^A(\sigma,\tau)$ obeys a linearized perturbation around
$q^A(\sigma,\tau)$. $\xi^A(\sigma,\tau)$ is a solution of second perturbative
order around $q^A(\sigma,\tau)$. Higher order perturbations can be considered
systematically. A physically appealing starting solution is the center of
mass motion of the string, $q^A(\tau)$, that is, the point particle (geodesic)
motion. The world sheet time variable appears here naturally identified with
the proper time of the center of mass trajectory. The space time geometry is
treated {\it exactly}, and the string fluctuations around $q^A$ are treated
as perturbations. Even at the level of the zeroth order solution, gravitational
effects including those of the singularities of the geometry are fully taken
into account. This expansion corresponds to low energy excitations of the
string as compared with the energy associated to the geometry. This corresponds
to an expansion in powers of $(\alpha')^{1/2}$. Since $\alpha'=(l_{Planck})^2$,
the expansion parameter turns out to be the dimensionless constant
$$g=l_{Planck}/R_c=1/(l_{Planck}M)~~,\eqno(i2)$$
where $R_c$ characterizes the spacetime curvature and $M$ is its associated
mass (the black hole mass, or the mass of a closed universe in cosmological
backgrounds). The expansion is well suited to describe strings in strong
gravitational regimes (in most of the interesting situations one has clearly
$g\ll1$). The constraint equations are also expanded in perturbations. The
classical $({\rm mass})^2$ of the string is defined through the center of
mass motion (or Hamilton - Jacobi equation). The conformal generators (or
world - sheet two dimensional energy - momentum tensor) are bilinear in the
fields $\eta^A(\sigma,\tau)$. [If this method is applied to flat spacetime,
the zeroth order plus the first order fluctuations provide the exact solution
of the string equations].

This method was first applied to cosmological (De Sitter) spacetimes. One of
the results was that for large enough Hubble constant, the frequency of the
lower string modes, i.e. those with $|n|<\alpha' mH$, ($\alpha'$ being the
string tension and $m$ its mass), becomes imaginary. This was further analysed
\refto{sv90,gsv} as the onset of a physical instability, in which the
proper string size starts to grow (precisely like the expansion factor of the
universe). The string modes couple with the background geometry in such a way
that the string inflates with the universe itself. The same happens for
strings in singular gravitational plane waves\refto{vs91,vms} (see
also ref[\cite{LS92b}]), and the
results of paper \cite{LS93} show that this is a generic feature of strings
near spacetime singularities.

In black hole spacetimes, such unstable features have been explored
in \ref{LS93}.
The string dynamics in black hole spacetimes is much more complicated to
solve (even asymptotically and approximately). In ref [\cite{vs88}], the
study of string dynamics in a Schwarzschild black hole was started and the
scattering problem was studied for large impact parameters. Stable oscillatory
behavior of the string was found for the transversal (angular) components;
scattering amplitudes, cross section and particle transmutation process were
described, and explicitly computed in an expansion in $(R_s/b)^{D-3}$, $R_s$
being the Schwarzschild radius and $b$ the impact parameter. The aim of
paper \cite{LS93}, was to find, and then to describe,
the {\it unstable sector} of strings in black hole backgrounds. By unstable
behavior, we mean here the following characteristic features: non oscillatory
behavior in time, or the emergence of imaginary frequencies for some modes,
accompanied of an infinite stretching of the proper string length. In addition,
the spatial coordinates (some of its components) can become unbounded.
Stable string behavior means the usual oscillatory propagation with real
frequencies, (and the usual mode - particle interpretation), the fact that
the proper string size does not blow up, and that the string modes remain
well - behaved.

We express the first order string fluctuations $\eta^\mu$, $(\mu=0,....,D-1)$
in $D$ - dimensional Reissner - Nordstr\"om - De Sitter spacetime, as a
Schr\"odinger type equation for the amplitudes $\Xi^\mu=q^R\eta^\mu$,
$q^R$ being the radial center of mass coordinate. We find the asymptotic
behavior of the longitudinal and transverse string coordinates ($\Xi^+,
\Xi^-,\Xi^i$) with $i=2,.....,D-1$, at the spatial infinity, near the
horizon and near the spacetime singularity. $+$ and $-$ stand for the
longitudinal (temporal and radial) components respectively, and $i$ for the
transverse (angular) ones. We analyse first a head - on collision
(angular momentum $L=0$), that is, a radial infall of the string towards
the black hole. Then, we analyse the full $L\not=0$ situation. We consider
Schwarzschild, Reissner - Nordstr\"om and De Sitter spacetimes (described
here in static coordinates which allow a better comparison among the three
cases). In all the situations (with and without angular momentum) and for the
three cases we find the following results:

The time component $\Xi^+$ is {\it  {always stable}} in the three
regions (near infinity, the horizon and the singularity), and in
the three cases (black holes and De Sitter spacetimes).

The radial component $\Xi^-$ is {\it  {always unstable}} in the three
regions and in the three backgrounds. In the Schwarzschild case, the
instability condition for the radial modes - which develop imaginary
frequencies near the horizon - can be expressed as
$$n<{\alpha' m \sqrt{D-3}\over R_s}\left[D-2-\left({D-3\over2}\right)
{m^2\over E^2}\right]^2~,\eqno(i3)$$
where $\alpha'$, $m$ and $E$ are the string tension, string mass and energy
respectively. The quantity within the square brackets is always positive,
thus the lower modes develop imaginary frequencies when the typical string size
$\alpha' m \sqrt{D-3}$ is larger than the horizon radius. Notice the
similarity with the instability condition in De Sitter space,
$n<\alpha' m/r_H$, $r_H$ being the horizon radius.

In the Schwarzschild black hole, the transverse modes $\Xi^i$ are stable
(well behaved) everywhere including the spacetime singularity at $q^R=0$.
In the Reissner - Nordstr\"om (R-N) black hole, the transverse modes
$\Xi^i$ are stable at infinity and outside the horizon. Imaginary
frequencies appear, however, inside a region from $r_-<q^R<r_+$ to $q^R\to0$,
where $r_\pm=M\pm \sqrt{M^2-Q^2}$, $M$ and $Q$ being the mass and charge of the
black hole respectively. For the extreme black hole $(Q=M)$, instabilities
do not appear. There is a critical value of the electric charge of a
Reissner - Nordstr\"om black hole, above which the string passing through
the horizon passes from unstable to stable regime. In the De Sitter
spacetime, the only stable mode is the temporal one $(\Xi^+)$. All the
spatial components exhibit instability, in agreement with the previous results
in the cosmological context\refto{vs87,sv90,gsv}. A summary of this analysis
is given in Table 1.

Imaginary frequencies in the transverse string coordinates $(\Xi^i)$
appear in the case in which the local gravity, i.e. $\partial_r a/2$, is
negative (that is, repulsive effects). Here,
$$a(r)=1-(R_s/r)^{D-3}+(\tilde Q^2/r^2)^{D-3}+
{\Lambda\over 3}r^2~,\eqno(i5)$$
where $\tilde Q^{2(D-3)}={8\pi GQ^2\over(D-2)(D-3)}$, and $\Lambda$ is the
cosmological constant. That is why the transverse modes $(\Xi^i)$ are
well behaved in the Schwarzschild case, and outside the Reissner - Nordstr\"om
event horizon.
But close to $q^R\to0$, $a'_{RN}<0$ (Reissner - Nordstr\"om has a repulsive
inner horizon),
and the gravitational effect of the charge overwhelms that of the mass; in
this case instabilities develop. In the Reissner - Nordstr\"om -
De Sitter spacetime, unstable string behavior appears far away from the
black hole where De Sitter solution dominates, and inside the black hole
where the Reissner - Nordstr\"om solution dominates.
For $M=0$ and $Q=0$, we recover the instability
criterion\refto{vs87,sv90} $\alpha'm\Lambda/6>1$ for large enough Hubble
constant (this is in agreement with the criterion given in ref[\cite{g92}].

We find that in the black hole spacetimes, the transversal first order
fluctuations $(\Xi^i)$ near the space time singularity $q^R=0$, obey
a Schr\"odinger type equation (with $\tau$ playing the role of a spatial
coordinate), with a potential $\gamma(\tau-\tau_0)^{-2}$,
(where $\tau_0$ is the proper time of arrival to the singularity at $q^R=0$).
 The dependence on
$D$ and $L$ is concentrated in the coefficient $\gamma$. Thus, the approach
to the black hole singularity is like the motion of a particle in a
potential $\gamma(\tau-\tau_0)^{-\beta}$, with $\beta=2$. And, then, like
the case $\beta=2$ of strings in singular gravitational waves\refto{vs91,vms}
(in which case the spacetime is simpler and the exact full string equations
become linear). Here $\gamma>0$ for strings in the Schwarzschild spacetime,
for which we have regular solutions $\Xi^i$ ; while $\gamma<0$ for
Reissner - Nordstr\"om,
that is, in the case we have a singular potential and an unbounded behavior
(negative powers in $(\tau-\tau_0)$) for  $\Xi^i_{RN}$.
The fact that the angular
coordinates $\Xi^i_{RN}$ become unbounded means that the string makes
infinite turns around the spacetime singularity and remains trapped by it.

For $(\tau-\tau_0)\to0$, the string is trapped by
the black hole singularity. In
Kruskal coordinates $\left(u_k(\sigma,\tau), v_k(\sigma,\tau)\right)$, for
the Schwarzschild black hole we find
$\lim_{(\tau-\tau_0)\to0}u_kv_k=\exp{2KC(\sigma)(\tau-\tau_0)^P}$,
 where $K=(D-3)/(2R_s)$
is the surface gravity, $P>0$ is a
determined coefficient that depends on the $D$ dimensions, and $C(\sigma)$
is determined by the initial state of the string. Thus $u_kv_k\to1$ for
$(\tau-\tau_0)\to0$. The proper spatial string length at fixed
$(\tau-\tau_0)\to0$ grows like $(\tau-\tau_0)^{-(D-1)P}$.

It must be noticed that in cosmological inflationary backgrounds, the
unstable behavior manifests itself as non - oscillatory
in $(\tau-\tau_0)$ (exponential
for $(\tau-\tau_0)\to\infty$, power - like
for $(\tau-\tau_0)\to0$); the string coordinates
$\eta^i$ are constant (i.e. functions of $\sigma$ only), while the proper
amplitudes $\Xi^i$ grow like the expansion factor of the universe.
In the black hole cases, and
more generally, in the presence of spacetime singularities, all the
characteristic features of string instability appear, but in addition the
spatial coordinates $\eta^i$ (or some of its components) become unbounded.
That is, not only the amplitudes $\Xi^i$ diverge, but also the string
coordinates $\eta^i$, what appears as a typical feature of strings near
the black hole singularities. A full description of the string behavior
near the black hole singularity will be reported elsewhere\refto{LS95}.

It is also worth noting that we have considered $m$, the mass of the string,
being different from zero, although to zeroth order the constraint equations
are satisfied by a null geodesic rather than by a timelike one. To obtain the
value of $m$ one has to consider second order perturbations, however, since
we were mainly concerned with the possibility of the string unstable behavior,
and not its detailed trajectory, first order analysis is enough.

\head{9. Formulation of the problem, Non-collinear collision and Discussion}

de Vega and S\'anchez\refto{vs87}
have obtained the equations of motion
of fundamental strings in curved
 backgrounds by expanding the fluctuations
of the string around a given particular
solution of the problem (for example, the center of mass
motion). For the case of a black hole background\refto{vs88}
with mass $M$, charge $Q$ and cosmological constant $\Lambda$,
$$ds^2=-a(r)(dX^0)^2+a^{-1}(r)dr^2+r^2d\Omega^2_{D-2}~,$$
$$a(r)=1-(R_s/r)^{D-3}+(\tilde Q^2/r^2)^{D-3}+
{\Lambda\over 3}r^2~,\eqno(1')$$
$$R_s^{D-3}={16\pi GM\over(D-2)\Omega_{D-1}}~~,~~\Omega_D={2\pi^{D/2}\over
\Gamma(D/2)}~~,~~\tilde Q^{2(D-3)}={8\pi GQ^2\over(D-2)(D-3)}~,$$
the equations of motion of the first order fluctuations read
$$\left[{d^2\over d\tau^2}+n^2\right] \eta_n^A(\tau)+
2A^A_B(\tau)\dot \eta^B_n(\tau)
+B^A_B(\tau)\eta^B_n(\tau)=0~.\eqno(c1)$$
Here we have expanded the first order string perturbations
in a Fourier transform
$$\eta^A(\sigma,\tau)=\sum_n e^{in\sigma}\eta_n^A(\tau)~.\eqno(c2)$$
$\eta^A$ being the vector
$$\eta=\pmatrix{\eta^0\cr \eta^*\cr \eta^i\cr}~~,~~i=2,3,.....,D-1\eqno(c3)$$
and $A^A_B(\tau)$ and $B^A_B(\tau)$ are the components of the following
matrices
$$A=\pmatrix{-{a'\dot q^R\over 2a} &
 -{\alpha'Ea'\over 2a}&{\bf O}\cr
           ~~&~~&~~\cr
	   -{\alpha'Ea'\over 2a} &
	   -{a'\dot q^R\over 2a}&
	   q^R\dot q^j\cr
	   ~~&~~&~~\cr
	   {\bf O} & {\dot q^i\over q^R}a&
	   {\dot q^R\over q^R}\delta^{ij}
	   +q^i\dot q^j\cr}~,\eqno(4)$$
$$B=\pmatrix{0&	-{\alpha'Ea''\dot q^R\over a}&{\bf O}\cr
             ~~&~~&~~\cr
             0& -S&{\bf O}\cr
             ~~&~~&~~\cr
	    {\bf O}&2\dot q^i\Bigl({a\over q^R}\dot{\Bigl)}
	      &\left({\alpha'L\over
	    (q^R)^2}\right)^2\delta^{ij}\cr}~,\eqno(5)$$
with $$S={a''\over 2a}\left[(q^R)^2+\alpha'^2E^2\right]-
\left({\alpha'L\over (q^R)^2}\right)^2a~,\eqno(c6)$$
Here the dot stands for $\partial/\partial\tau$ and
$'=\partial/\partial q^R$. The solution for the string
coordinates is given by
the expansion
$$X^\mu=q^\mu+\eta^\mu+\xi^\mu+....~~,~~\mu=0,1,....,D-1~.\eqno(7)$$
$q^\mu(\tau)$ being the center of mass
 coordinates (zeroth order solutions), which follow the
geodesics of the background space-time, i.e.
$$\dot q^0={\alpha'E\over a}~~,~~(\dot q^i)^2=
\left({\alpha'L\over (q^R)^2}\right)^2~~,~~
{(\dot q^R)^2\over\alpha'^2a}+{L^2\over(q^R)^2}+
m^2-{E^2\over a}=0~,\eqno(8)$$
where we have identified the proper
time of the geodesic with
the world-sheet $\tau$ - coordinate.

To study the equation of motion of the
first fluctuations, eq \(c1),
it is convenient to apply a transformation
 to the vector $\eta^A$.
Let us propose
$$\eta_n=G\Xi_n~,\eqno(9)$$
where the matrix $G$ is chosen to
 eliminate the term in the first
derivative in eq \(c1), i.e.
$$G=P\exp\{-\int^\tau A(\tau')d\tau'\}~,\eqno(10)$$
where $P$ is a constant normalizing matrix.
 Thus, eq \(c1) transforms into
$$\ddot\Xi_n+G^{-1}(n^2+B-\dot A-A^2)G\Xi_n=0~,\eqno(11)$$
which is a Scr\"odinger-type equation with $\tau$
playing the role of the spatial coordinate.

It is simple to analyse the first order
fluctuations in the transversal coordinates.
 In fact, for $i>2$, the matrices $A$ and $B$
  given by eqs \(4) and \(5) are diagonal,

$$A^{ij}={\dot q^R\over q^R}\delta^{ij}~~;~~
B^{ij}=\left({\alpha'L\over (q^R)^2}\right)
^2\delta^{ij}~,\eqno(102)$$
then the equations for the first order
fluctuations are
$$\ddot\Xi^i_n+(n^2+\left({\alpha'L\over
 (q^R)^2}\right)^2-\ddot q^R/q^R)\Xi^i
 _n=0~.\eqno(103)$$

By use of the geodesic equations \(8)
 we can rewrite eq \(103) as
$$\ddot\Xi^i_n+\left[n^2+{(\alpha')^2
\over 2}\left({a'(q^R)m^2\over q^R}+
2\left({L\over (q^R)^2}\right)^2\left(
1-a+{1\over2}a'q^R\right)\right)\right]
\Xi^i_n=0~.\eqno(104)$$

We can still re-write the above equations as

$$\ddot\Xi_n^\mu+[n^2+\lambda^\mu]\Xi_n^\mu=0~,
\eqno(59)$$
where
$$\Xi_n^\mu=(\Xi_n^+, \Xi_n^-, \Xi_n^i)~~;~~
\lambda^\mu=(\lambda^+, \lambda^-,
\lambda^i)~,\eqno(60)$$
and
$$\lambda^\pm=\lambda_\pm-n^2~~,~~\lambda^i
={(\alpha' m)^2\over 2}
{a'(q^R)\over q^R}~.\eqno(70)$$

Thus we are able to analyse now the particular
 cases we are interested in:

\subhead{9.1 Schwarzschild black hole}

Plugging eq \(1') into \(104)
 we obtain
$$\lambda^i_{Sch}={(\alpha')^2\over 2}R_s^
{D-3}(q^R)^{1-D}\left[(D-3)m^2+(D-1)\left(
{L\over q^R}\right)^2\right]~.\eqno(105)$$
We observe that for $D\geq4$, $\lambda^i_
{Sch}>0$. Thus producing stable first order
 fluctuations. In fact, when we study the
 time evolution we obtain
$$\lim_{q^R\to0}{\lambda^i_{Sch}}=
{2(D-1)\over(D+1)^2}(\tau-\tau_0)^{-2}~,\eqno(106)$$
where we have integrated the expression \(8),
$$\alpha'(\tau-\tau_0)=\int^{q^R}_{q^R_0}
{dq^R\over\sqrt{E^2-m^2a(q^R)-a\left({L\over
 q^R}\right)^2}}~, \eqno(107')$$
in order to obtain the behavior
$$q^R_S(\tau)\to\left[{(D+1)\over2}\alpha'R_s
^{(D-3)/2}L(\tau-\tau_0)\right]^{2/(D+1)}.
\eqno(107)$$
Eq \(105) has, then, a power-like solution
in the limit $q^R\to0$,
$$\Xi_{n, Sch}^i(\tau)\sim \left[n(\tau-\tau_0)
\right]^P~~;~~P^i_{Sch}={1\over2}\pm
\sqrt{{1\over4}-{2(D-1)
\over(D+1)^2}}~.\eqno(108)$$
That vanishes for $(\tau-\tau_0)\to0$.

\subhead{9.2 De Sitter spacetime}

This case is very interesting because replacing
 expressions \(1') into the first
 order fluctuations equations with $L\neq0$,
  eq \(104), we obtain that the $L$
  dependence disappears. Giving, in fact,
   the equation
$$\ddot\Xi^i_n+\left[n^2-{(\alpha'm)^
2\over6}\Lambda\right]
\Xi^i_n=0~.\eqno(109)$$
The solution to this equation are exponentials
 in the proper time, $\tau$,
suggesting the occurrence of instabilities.

The fact that the solutions
should be $L$-independent,
could have been guessed from the symmetries
 of the De Sitter space. This metric has
  not preferred point to refer the angular
 moment to as in the black hole cases (there
  is not singularity at $r=0$). This allows
  us to say that the components $\Xi^-$,
   $\Xi^+$ and $\Xi^i$ will behave as the
   ones already studied for the case $L=0$.

\subhead{9.3 Reissner-Nordstr\"om black hole}

{}From the metric coefficients \(1') and its
first derivative we find that eq \(104) reads
$$\ddot\Xi^i_n+\eqno(110)$$
$$+\left\{n^2+{(\alpha')^2\over2}
\left[{(D-3) m^2
\over (q^R)^2x}(1-{2\beta\over x})+{L^2\over
x(q^R)^4}\left((D-1)-{2\beta\over x}(D-2)
 \right)\right]\right\}\Xi^i_n=0~.$$
We observe that as $q^R\to\infty$, the term
proportional to $L^2$ vanish faster than the
other terms. Thus, in this limit we recover
the $L=0$ results.

As we approach to the black hole, and arrive
to the horizon, we have for the Schwarz\-schild
 black hole
$$\lim_{q^R\to q^R_H}{\lambda^i_{H,Sch}}=
{(\alpha')^2\over2}\left[
{(D-3)m^2\over R_s^2}+(D-1){L^2\over R_s^4}
\right]~,\eqno(111)$$
which is definite positive, and does not produce
 instabilities.

For the extreme Reissner-Nordstr\"om black
hole we have,
$$\lim_{q^R\to q^R_H}{\lambda^i_{H,ERN}}=
(\alpha')^22^{4/(D-3)}{L^2\over R_s^4}~.\eqno(112)$$
Again, it is always positive. Thus, we see
 that the instabilities do not appear yet.
  The time-dependence of the solutions
  close to the horizon will be oscillatory
   with the squared frequency given by
   ${\lambda^i_{H,ERN}}$.

However, the picture changes when we go
closer to the singularity. For the Reissner-Nordstr\"om
black hole, when $q^R\to0$, we have
$$\lim_{q^R\to0}{\lambda^i_{0,RN}}=
-(\alpha')^2
{(D-2)\beta\over x^2}{L^2\over (q^R)^4}~.\eqno(113)$$
As this squared frequency takes negative
values  allows the possibility that
instabilities develop in
the string transversal coordinates.
In order to find the time dependence in the
coordinates we integrate first the
center of mass motion, eq \(8). Then,
as $q^R\to0$, we obtain
 $$\alpha'(\tau-\tau_0)\sim{x\over L\sqrt{
 \beta}}{(q^R)^2\over(D-1)}~.\eqno(114)$$
Plugging this expression into eq \(113),
 we have
$$\lim_{q^R\to0}{\lambda^i_{0,RN}}\to-
{(D-2)\over(D-1)^2}(\tau-\tau_0)^{-2}~. \eqno(115)$$
And the solution of the first order
 fluctuations is again a power-like one,
$$\Xi_{n, RN}^i(\tau)\sim \left[n(\tau-
\tau_0)\right]^P~~;~~P^i_{RN}={1\over2}
\pm\sqrt{{1\over4}+{(D-2)
\over(D-1)^2}}~.\eqno(116)$$
We have here that the solution with minus
sign in front of the square root produces
an unbounded solution as $(\tau-\tau_0)\to0$,
 thus, suggesting the existence of
 instabilities.

It is worth to remark that the solutions
\(116) and \(108) for the time dependence
 of the transversal coordinates for the
 Reissner-Nordstr\"om and Schwarzschild
 black holes respectively, are independent
 of $L$. They are, although, different from
 those of the case $L=0$ (see \ref{LS93}).
This is so because even if the $L$
dependence cancels out from the final
equations \(116) and (108), the approach
to the singularity, $q^R\to0$, is
different if $L\not=0$, thus producing
different final coefficients.

Another interesting feature of the equations
for the transversal first order modes, is
that for the black hole cases (Schwarzschild
or Reissner-Nordstr\"om; orbit of the
string center of mass with or without
angular momentum), the time dependence
of $\lambda^i$ appears to be $(\tau-\tau_0)^{-2}$
as $q^R\to0$. The behavior of $\lambda^i$
as a function of $q^R$ is different for each case,
but the $\tau$ - dependence of the orbit in each case
exactly compensates for such difference.

Thus, for the linearized string fluctuations,
the approach to the black hole singularity
corresponds to the case $\beta=2$ of the motion of a particle
in a potential $\gamma(\tau-\tau_0)^{-\beta}$. This is like
the case of strings in singular
plane - wave backgrounds.\refto{vs91,vms} In fact, the linearized
first order string fluctuations produce a one -
dimensional Schr\"odinger equation, with
$\tau$ playing the role of a spatial coordinate.
The potential term in eq \(104), can be
written fully $\tau$ - dependent, as we
have seen, by plugging into it the center
of mass trajectory, $q^R(\tau)$.(In the case of gravitational
plane waves, the spacetime is simpler than  in the black hole
one, and the exact full string equations become linear).

The solution of eq \(104) with a potential
proportional to $\gamma(\tau-\tau_0)^{-2}$
can be given in terms of Bessel functions,
$$\Xi_{n}^i(\tau)=\sqrt{(\tau-\tau_0)}\left
\{V_{n}^iJ_\nu[n(\tau-\tau_0)]+
W_{n}^iJ_{-\nu}[n(\tau-\tau_0)]\right\}~.\eqno(117)$$
Where $V_{n}^i$ and $W_{n}^i$ are arbitrary
constants coefficients and
$$\nu=\sqrt{{1\over4}-\gamma}~~~~{\rm i.e.}~,~~\nu=P^\mu-1/2~,\eqno(118)$$
Where $P^\mu(D,\beta)$'s are defined in Section IV of \ref{LS93}.

For $\gamma<0$ we have
Bessel functions (those with negative index)
with a divergent behavior as $(\tau-\tau_0)\to0$,
indicating the existence of
string instabilities.
We would also to stress that for black
holes, what determinates the possibility
of instabilities is not the type of singularity
(Reissner-Nordstr\"om or Schwarzschild),
nor how it is approached ($L=0$ or $L\not=0$),
because we have seen that the time dependence
of the potential is always $(\tau-\tau_0)^{-2}$,
 but the sign of the coefficient $\gamma$ in
front of it, i.e. the attractive character of the potential
$(\tau-\tau_0)^{-2}$. Thus, we can
conclude that whenever we have big enough
repulsive effects in a gravitational background,
instabilities in the
propagation of strings on the
spacetime background will appear.
The coefficient $\gamma$ is given by
$$\gamma_{Sch}=\cases{{2(D-3)\over(D-1)^2}~~,~~L=0\cr
\cr
                      {2(D-1)\over(D+1)^2}~~,~~L\not=0\cr}$$
for the Schwarzschild black hole, and
$$\gamma_{RN}=\cases{-{2(D-3)\over(D-2)^2}~~,~~L=0\cr
\cr
                      {-2(D-2)\over(D-1)^2}~~,~~L\not=0\cr}$$
for the Reissner-Nordstr\"om black hole.

Near the spacetime singularity, the dependence on the $D$ spacetime
dimensions is concentrated in $\gamma$. Notice the attractive singular
character of the potential $\gamma(\tau-\tau_0)^{-2}$, for the
Reissner-Nordstr\"om black hole, in agreement with the singular behavior
of the string near $q^R=0$; while for the Schwarzschild black hole
$\gamma$ is positive, and the string solutions $\Xi^i$ are well
behaved there.

The approach to the black hole singularity is better analysed in
terms of the Kruskal coordinates $(u_k, v_k)$
$$u_k=e^{Ku_{Sch}}~~~,~~~v_k=e^{Kv_{Sch}}$$
$u$ and $v$ being null coordinates, and $K$ the surface gravity of the
black hole $(K=(D-3)/(2R_s)$ for Schwarzschild).
We also have $(C^\pm(\sigma)$ being coefficients determined by the initial
state of the string)
$$u_k(\sigma,\tau)=\exp{K\left[C^+(\sigma)(\tau-\tau_0)^{P^+}-
C^-(\sigma)(\tau-\tau_0)^{-|P^-|}\right]}$$
$$v_k(\sigma,\tau)=\exp{K\left[C^+(\sigma)(\tau-\tau_0)^{P^+}+
C^-(\sigma)(\tau-\tau_0)^{-|P^-|}\right]}\eqno(u1)$$
and thus, $u_kv_k\to1$ for $(\tau-\tau_0)\to0$.
 That is, for $(\tau-\tau_0)\to0$, the
string approaches the spacetime singularity $u_kv_k=1$, and it is
trapped by it. The proper spatial length element of the string at
fixed $(\tau-\tau_0)\to0$, between $(\sigma,\tau)$ and $(\sigma+d\sigma,\tau)$,
stretches infinitely as
$$ds^2_{(\tau-\tau_0)\to0}\to \left(C^{-}(\sigma)\right)'^2d\sigma^2
(\tau-\tau_0)^{-(D-1)|P^-|}~,\eqno(u2)$$
where $P^-$ is given by eq (71) of \Ref{LS93}.
Here, $\tau_0$ is the (finite) proper
falling time of the string into the black hole singularity.

The fact that the angular coordinates $\Xi^i$ become un\-bounded in the
Reissner - Nordstr\"om case, means that the string makes infinite turns
around the spacetime singularity and remains trapped by it.

It is also worth noting that we have considered $m$, the mass of the string
being different from zero although to zeroth order the constraint equations
are satisfied by a null geodesic rather than a timelike one. To obtain the
value of $m$ one has to consider second order perturbations, however, since
we were mainly concerned with the possibility of the string unstable behavior,
first order analysis is enough.
The same conclusions can be drawn for
the quantum propagation of
strings. The $\tau$ dependence is the same
because this is formally described
by a Schr\"odinger equation
with a potential $\gamma(\tau-\tau_0)^{-2}$,
the coefficients of the solutions being quantum operators instead of
C - numbers. The $\tau$ evolution of the string near the black hole
singularity is fully determined by the spacetime geometry, while the
$\sigma$ - dependence (contained in the overall coefficients) is fixed
by the state of the string.

For the modes $\Xi^-$, $\Xi^+$ and $\Xi^i$
we can conclude that they should behave as
in the case $L=0$ far from the black hole,
where the influence of the angular momentum
vanishes. Then, $\Xi^+$ and $\Xi^i$ will
oscillate with bounded amplitude outside the horizon while
$\Xi^-$ will present an unbounded behavior.
The approach to the singularity with
$L\not=0$ should not change qualitatively from the picture for $L=0$.
The analysis can be also made in terms of
the geodesic orbit followed by the center
of mass of the string. For a given energy
$E$, there is a critical impact parameter
$b_c$ which determines whether the string
will fall or not into the black hole. From
our results here for $L=0$ (see
Table 1), and for $L\not=0$, we can draw
the following picture: For large impact
parameters
$b>b_c$, the transversal, $\Xi^i$, and temporal, $\Xi^+$, modes
will be stable, while the radial modes, $\Xi^-$, begin to
suffer instabilities. For small impact
parameters, $b<b_c$, the string will fall
into the black hole and, for a Reissner-
Nordstr\"om background, also the transversal
 coordinates suffer instabilities.

It can be noticed that in cosmological inflationary backgrounds,
for which unstable string behavior appears when $(\tau-\tau_0)\to0$, the
string coordinates $\eta^i$, remain bounded. In the black hole
cases, all the characteristic features of string instabilities
appear, but, in addition, the string coordinates $\eta^i$ become
unbounded near the $r=0$ singularity. This happens to be a typical
behavior of strings near spacetime singularities, describing the fact
that the string is trapped by it.

\head{10. Acknowledgements}
\cee ~\guita ~C.O.L.
also thanks the DEMIRM - Observatoire de Paris, where part of this
work was done, for kind hospitality and working facilities.

\vfill\eject
\bigskip\baselineskip=10pt
\bigskip\noindent
{\bf Table 1} Regimes of string stability in black hole and De Sitter
spacetimes:  Here stable means well behaved string fluctuations
and the usual oscillatory behavior with real frequencies. Unstable
behavior corresponds to unbounded amplitudes $(\eta^\pm,\eta^i)$
with the emergence of non oscillatory behavior or imaginary
frequencies, accompanied by the infinite string stretching of the
proper string length. $\eta^+$,$\eta^-$ and $\eta^i$, $(i=2,...,
D-1)$, are the temporal, radial and angular (or transverse) string
components respectively.

\bigskip
\bigskip
\bigskip
\def\smallrule{\makeline{\II\noline\I\thinline\thinline\thinline\thinline}}

\table{15truecm}{
      \====
      \[  Region        |
          Mode          |
	  Schwarzschild |
	  Reissner-Nordstr\"om |
	  De Sitter    \]
      \====
      \[              |
          $\eta^{+}$   |
	  stable      |
	  stable      |
	  stable     \]
       \smallrule
       \[ $q^R \to 0$  |
          $\eta^{-}$    |
	  unstable   |
	  unstable   |
	  unstable  \]
       \smallrule
       \[            |
          $\eta^{i}$  |
	  unstable   |
	  unstable   |
	  unstable  \]
	\====
       \[            |
	  $\eta^{+}$  |
	  stable     |
	  stable     |
	  stable    \]
       \smallrule
       \[ $q^R \to q_H^R$   |
          $\eta^{-}$         |
          unstable          |
	  unstable / stable |
          unstable         \]
       \smallrule
       \[            |
          $\eta^{i}$  |
	  stable     |
	  stable     |
	  unstable  \]
       \====
       \[            |
          $\eta^{+}$  |
	  stable     |
	  stable     |
	  stable    \]
       \smallrule
       \[ $q^R \to \infty$  |
          $\eta^{-}$  |
          unstable   |
	  unstable   |
	  unstable  \]
       \smallrule
       \[            |
          $\eta^{i}$  |
	  stable     |
	  stable     |
	  unstable  \]
       \====  }

\vfill\eject
\references
\bigskip
\baselineskip=14pt

\refis{HS94}K. Hayashi and T. Samura, gr-qc/9404027; ibid hep-th/940513.

\refis{AB94}P. C. Aichelburg and H. Balasin, gr-qc/9407018.

\refis{LS89b}C.O.Lousto and N.S\'anchez, \plb 232,462,1989.

\refis{DH85}T.Dray and G.'t Hooft, \npb 253,173,1985.

\refis{LS89a}C.O.Lousto and N.S\'anchez, \plb 220,55,1989.

\refis{LS90}C.O.Lousto and N.S\'anchez, \ijmpa 5,915,1990.

\refis{dVS89b}H.J.de Vega and N.S\'anchez, \npb 317,731,1989.

\refis{dVS89a}H.J.de Vega and N.S\'anchez, \npb 317,706,1989.

\refis{AS71}P.C.Aichelburg and R.U.Sexl, \grg 2,303,1971.

\refis{10} V.Ferrari and P.Pendenza, \grg 22,1105,1990.

\refis{2} C.W.Misner, K.S.Thorne and J.A.Wheeler, {\it Gravitation}, (W.H.
Freeman and Co., San Francisco, 1973).

\refis{H87}G.'t Hooft, \plb 198,61,1987.

\refis{LS92b} C. O. Lousto and N. S\'anchez,\prd 46,4520,1992.

\refis{LS93} C. O. Lousto and N. S\'anchez,\prd 47,4498,1993.

\refis{LS91} C.O.Lousto and N.S\'anchez, \npb 355,231,1991.

\refis{H88} G.'t Hooft, \npb 304,867,1988.

\refis{GV91} J.Garriga and E.Verdaguer, \prd 43,391,1991.

\refis{ACV88}D.Amati, M.Ciafaloni and G. Veneziano, \ijmpa 3,1615,1988.

\refis{GM88}D.Gross and P.Mende, \npb 303,407,1988.

\refis{AK88}D.Amati and C.Klim\v cik, \plb 210,92,1988.

\refis{V88} G.Veneziano, "Strings and Gravitation", talk presented at the {\it
5th Marcel Grossmann meeting} (Perth, Aug. 1988), D.G.Blair and M.J.Buckinghan
eds. (World Sci., Singapore, 1989), p. 173.

\refis{dVS90}  H.J. de Vega and N.S\' anchez, \plb 244,215,1990. Ibid
\prl 65,1517,1990.

\refis {23} G.T.Horowitz and A.R.Steif, \prl 64,260,1990. Ibid
\prd 42,1950,1990.

\refis{CdV91} M.E.V.Costa and H.J. de Vega, \annp 211,223,1991. Ibid,
\annp 211,235,1991.

\refis{LS92a}C.O.Lousto and N.S\'anchez, \npb 383,377,1992.

\refis{17} G. Veneziano, {\it DST Workshop on String Theory}, H. S. Mani and
R. Ramachandran eds. (World Sci. Publ. Co. 1988) p. 1.

\refis{19} A. Erdelyi et al., {\it Tables of Integral Transforms}, Vol.II,
 (McGraw-Hill, New York, 1954).

\refis{BN93} H. Balasin and H. Nachbagauer, \cqg 10,2271,1993.
Ibid gr-qc/9312028; ibid gr-qc/9405053.

\refis{vs88} H.J.de Vega and N.S\'anchez, \npb 309,552,1988.
Ibid  \npb 309,577,1988.

\refis{vs87} H.J.de Vega and N.S\'anchez, \plb 197,320,1987.

\refis{sv90} N.S\'anchez and G.Veneziano, \npb 333,253,1990.

\refis{gsv} M.Gasperini, N.S\'anchez and G.Veneziano, \ijmpa 6,3853,1991.
Ibid \npb 364,365,1991.

\refis{LS95} C.O.Lousto and N.S\'anchez, in preparation.\par

\refis{vs91} H.J.de Vega and N.S\'anchez, \prd 45,2783,1992.

\refis{vms} H.J.de Vega, M.Ram\'on-Medrano and N.S\'anchez, \npb 374,425,1992.

\refis{g92} M.Gasperini, \grg 24,219,1992.

\refis{ks94} K.Sfetsos, hep-th/9408169.

\endreferences
\end